\documentclass{aa}
\usepackage{graphicx}
\usepackage{natbib}
\usepackage{txfonts}
\makeatletter
\newcommand{\erf}{\mathop {\operator@font erf}\nolimits}
\makeatother
\def\dy{\displaystyle}

\def\hd{\object{HD\,141569\,A}}
\def\hd181{\object{HD\,181327}}
\def\bp{\object{$\beta$\,Pictoris}}

\def\mic{\object{AU\,Mic}}
\def\d{\mathrm d}

\newcommand{\uma}[1]{^{\mathrm{#1}}}
\newcommand{\dma}[1]{_{\mathrm{#1}}}
\newcommand{\bpr}{\beta\dma{pr}}
\newcommand{\bprq}{\beta\dma{pr,q}}
\newcommand{\bprf}{\beta\dma{pr,f}}
\newcommand{\bsw}{\beta\dma{sw}}
\newcommand{\bpreff}{\beta\dma{pr,eff}}
\newcommand{\pdx}[2]{#1\times10^{#2}}
\DeclareMathAlphabet{\mathpzc}{OT1}{pzc}{m}{it}
\begin{document}
\title{On the AU\,Mic Debris Disk}
\subtitle{Density Profiles, Grain Properties and Dust Dynamics}
\author{J.-C. Augereau\inst{1,2} \and H. Beust\inst{1}}
\institute{
Laboratoire d'Astrophysique de l'Observatoire de Grenoble, B.P. 53,
38041 Grenoble Cedex 9, France
\and
Leiden Observatory, PO Box 9513, 2300 RA Leiden, The Netherlands}
\offprints{J.C. Augereau} 
\mail{augereau@obs.ujf-grenoble.fr}
\date{\today} 
\titlerunning{The AU\,Mic debris disk}
\authorrunning{J.-C. Augereau \& H. Beust}
%%%%%%%%%%%%%%%%%%%%%%%%%%%%%%%%%%%%%%%%%%%%%%%%%%%%%%%%%%%%%%%%%%%%%%%%%%
\abstract
%
% Context: -------------------
{
\mic\ is a young M-type star surrounded by an edge-on optically thin
debris disk that shares many common observational properties
with the disk around \bp.
In particular, the scattered light surface brightness profile falls off as
$\sim r^{-5}$ outside 120\,AU for \bp\ and 35\,AU
for \mic. In both cases, the disk color
raises with increasing distance beyond these reference radii.
}
%
% Aims: ------------------------
{
We present the first comprehensive analysis
of the \mic\ disk properties since the system was discovered
by Kalas et al.\,(2004). We explore whether
the dynamical model, successful to reproduce the $\beta$\,Pic brightness profile
(e.g. Augereau et al.\,2001), could apply to \mic.
}
%
% Methods:
{
We calculate the surface density profile of the \mic\ disk by performing the
inversion of the near-IR and visible scattered light brightness profiles
measured by Liu\,(2004a) and Krist et al.\,(2005), respectively.
We discuss the grain properties by analysing the blue color of the
disk in the visible (Krist et al.\,2005) and by fitting the disk spectral
energy distribution. We finally evaluate the radiation and wind
forces on the grains. The impact of the recurrent X-ray and UV-flares
on the dust dynamics is also discussed.
}
%
% Results:
{
We show that irrespective of the mean scattering asymmetry factor of the grains,
most of the emission arises from an asymmetric, collisionally-dominated
region that peaks close to the surface brightness break around $35\,$AU.
The elementary scatterers at visible wavelengths are found to be sub-micronic,
but the inferred size distribution underestimates the amount of large
grains, resulting in too low sub-millimeter emissions compared to the observations.
{From our inversion procedure, we find that the V- to H-band scattering
cross sections ratio increases outside $40$\,AU, in line
with the observed color gradient of the disk.
This behaviour is expected if the
grains have not been produced locally but placed in orbits of high
eccentricity by a size-dependent pressure force, resulting in a paucity
of large grains beyond the outer edge of the parent bodies disk.}
Because of the low luminosity of \mic, radiation pressure is inefficient
to diffuse the smallest grains in the outer disk,
even when the flares are taken into account.
Conversely, we show that a standard, solar-like stellar wind generates a
pressure force onto the dust particles that behaves much like a radiation
pressure force. With an assumed $\dot{M} \simeq \pdx{3}{2}\dot{M}_{\odot}$,
the wind pressure overcomes
the radiation pressure and this effect is enhanced by the stellar
flares. This greatly contributes to populating the extended \mic\ debris disk
and explains the similarity between the \bp\ and \mic\ brightness
profiles. In both cases, the color gradient beyond $120\,$AU for \bp\
and $35\,$AU for \mic, is believed to be a direct consequence of the dust
dynamics.
}
%
% Conclusion:
{}
\keywords{Stars: circumstellar matter -- Stars individual: \mic --
Stars: flare -- Planetary systems: formation
-- Scattering}
\maketitle
%
%%%%%%%%%%%%%%%%%%%%%%%%%%%%%%%%%%%%%%%%%%%%%%%%%%%%%%%%%%%%%%%%%%%%%%%%%%
\section{Introduction}
\label{intro}
Recent visible and near-infrared coronagraphic imaging of the nearby
M-type star \mic\ revealed its geometrically and optically thin
circumstellar dust disk
\citep{kal04,liu04a,kri05,met05,fit05}. The disk, seen almost perfectly edge-on,
 extends over several hundreds of AU diameter. The disk
morphology presents interesting similarities
with the well-studied disk around the young A5V main sequence
star \bp: both disks are seen edge-on, imaging has evidenced
radial and vertical asymmetries, and importantly the radial surface
brightness profiles have similar shapes. Both dust disks are furthermore
thought to be made of short-lived collisional debris and the gas
to be largely dissipated \citep[][]{liu04b,bra04,rob05}.
\mic\ and \bp\ would moreover be co-eval since they would
belong to the same nearby stellar association aged $12^{+8}_{-4}$\,Myr
\citep{bar99,zuc01}.

Do these similarities between the two debris disks point toward similar ongoing
dynamical processes? The vertical asymmetries of the \bp\ disk seen
in scattered light have been modeled assuming the presence of a planet
on an inclined orbit with respect to the disk midplane
\citep{mou97,aug01}. The \mic\ disk appears slightly vertically
distorted within 50\,AU radius
but current observations do not readily show large-scale
vertical asymmetries similar to those observed in the \bp\ disk.
Hypothetic planetary companion(s) around \mic\ should therefore be on
low inclination orbit(s) with respect to the disk midplane.
The \mic\ disk also shows small-scale structures appearing as local
brightness enhancements
not present on each side of the disk \citep{liu04a,kri05}.
Such small-scale side-to-side asymmetries have also been 
reported for \bp\ \citep{pan97,tel05}
and could correspond to local enhancements of the collision rate
due to planetesimals trapped with a planet \citep[e.g.][]{roq94,tel05}.

More intriguing is the similarity between the midplane
surface brightness profiles of the \mic\ and \bp\ disks in
scattered light. The observed profiles both decrease almost linearly
with increasing the distance from the star then
drop with power law indexes  between $-4.4$ and $-5.5$.
The break between the two power-law regimes is observed around
35\,AU for \mic\ and 120\,AU for \bp. According to \citet{aug01},
the surface brightness break around 120\,AU in the \bp\ disk corresponds to the outer
edge of the collisional planetesimal ring that produces the observed dust
particles. The stellar radiation pressure of the central A5V star places the
smallest bound dust particles on orbits of high eccentricity with the same
periastron as their parent bodies. These grains spend a large fraction
of their time close to apoastron distributing this way the dust particles
over a wide range of distances far from the place they have been produced.
This results in a disk that extends over hundreds of AU with a generic
scattered light surface brightness profile further the 120\,AU break
and proportional to $r^{-5}$ as observed \citep[][]{lec96,aug01,the05}.
Because of the drag gas can induce on the dust particles, this model
requires a low gas density to keep the scattered light profile
consistent with the observations. \citet{the05} use this effect to place
an upper limit on the gas density in the \bp\ disk.

The previous model, successful to explain the scattered light brightness
profile of the \bp\ disk, does not \emph{a priori}
apply to the \mic\ disk because of the
low stellar luminosity of the central M-type star.
The stellar radiation pressure on the grains is expected to be small
and the steep slope of the scattered light profile further
35\,AU would then reflect the profile of the parent bodies distribution
rather than the effect of the radiation pressure like for \bp\ \citep{kal04}.
In this scenario, the similarities between the \mic\ and \bp\ surface brightness
profiles would just be coincidental.
This conclusion nevertheless depends on uncertain parameters. Radiation
pressure depends on the grains properties (composition, size) which are poorly
constrained. \mic\ is also a flare star with UV and X-Ray rises that could
periodically increase the radiation pressure force on the grains. Alternatively,
a hypothetical stellar wind could create a corpuscular pressure
force that would help diffusing the dust particles at large distances.
Finally, the collisional activity of the disk
has not been precisely determined and requires the disk surface density
to be properly estimated.

In this paper, we invert HST visible and ground-based
near-infrared surface brightness observations to calculate surface
density profiles. 
We derive in Sect.~\ref{inversion} the integral equation that links the
observed profiles to the physical quantities we want to recover, and in
Sect.~\ref{method} we detail our numerical approach.
We then discuss the shape of the disk surface density as a function
of the anisotropic scattering properties of the grains (Sect.~\ref{aumic}).
In Sect.~\ref{diskprop} we constrain the grain size distribution based
on the blue color of the disk in the visible \citep{kri05} and we derive
some disk properties (mass, optical thickness).
Finally we argue in Sect.~\ref{rad} and \ref{wind} that the observed
$r^{-5}$ surface brightness profile further
35\,AU in scattered light is not coincidental but instead reflects, like in the \bp\ 
disk, the ongoing dynamics of the observed dust grains. Our scenario
involves both the observed stellar flares and the presence of a wind drag force. 
%
%----------------------------------------------------------------------------
\section{Surface brightness profile}
%----------------------------------------------------------------------------
\label{inversion}
%
%
%----------------------------------------------------------------------------
\subsection{Notations and assumptions}
\label{hyp}
\begin{figure}[tp]
\centering
\includegraphics[angle=-90,origin=bl,width=0.95\columnwidth]{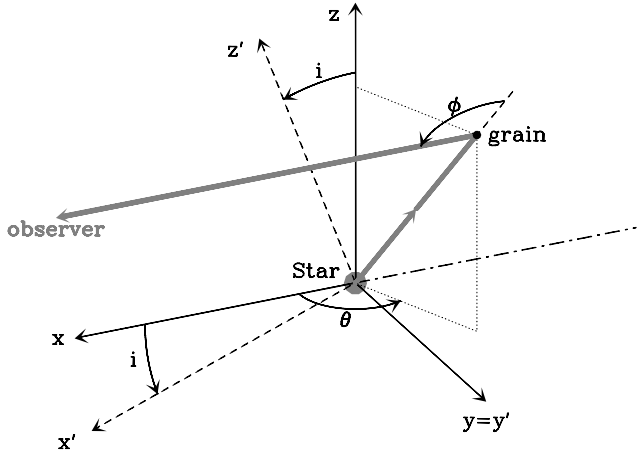}
\caption{Notations used in this study (see Sect.\,\ref{hyp})}
\label{sketch}
\end{figure}

We note $(x,y,z)$ the observer's line of sight Cartesian frame with
the star at the origin and with the $x$-axis pointing to the observer
(Fig.~\ref{sketch}).
We consider a disk rotated around the $y$-axis by an angle $i$
with $i=0$ corresponding to an edge-on disk. We assume 
$-\pi/2< i < \pi/2$ to have $\cos i>0$. 
We note $(x',y',z')$ the Cartesian frame attached to the disk
with the star at the origin and with the $z'$-axis perpendicular to the
disk midplane. We let $y'=y$ which implies that the $y'$-axis corresponds
to the apparent major axis of the disk when $i\ne 0$.
We furthermore define $(r,\theta,z)$
and $(r',\theta',z')$ the cylindrical coordinates
in the line of sight and disk frames, respectively.
We let $r'\dma{max}$ the maximal radius of the disk and $d\dma{star}$
the stellar heliocentric distance.

The \mic\ disk is assumed to be optically thin in all directions
and at all wavelengths which implies that the grains are directly
illuminated by the central star. This assumption is justified by
the low fractional dust disk luminosity \citep[$L\dma{disk}/L_* \simeq
\pdx{6}{-4}$,][]{liu04b}, the low extinction of the star and 
the lack of central dark
lane along the disk midplane in the visible and at near-IR wavelengths
as regularly observed for optically thick edge-on disks.

To account for light anisotropic scattering by the grains,
we introduce the scattering angle $\phi$ 
with $\phi=0$ corresponding to forward scattering
(see Fig.~\ref{sketch}).
Since $r'\dma{max}\ll d\dma{star}$, we will assume in the following
that $\theta=-\phi$ for the grains with $z=0$.
%
%----------------------------------------------------------------------------
\subsection{Scattered light surface brightness profile}
The surface brightness of an optically thin disk
measured at a projected distance $y$ from the star
along the $y$-axis ({\it i.e.} at $z=0$) writes
\begin{eqnarray}
B(y) & = &\int_{-\infty}^{+\infty} \rho(r',\theta',z') \Phi(r,\phi) \d x 
\label{eq1}
\end{eqnarray}
where $\rho(r',\theta',z')$ is the disk number density and
$\Phi(r,\phi)$
is the total flux measured at Earth emitted by a grain at
position $(r,\theta$=$-\phi,z$=$0)$. This flux is a function of
the observing wavelength but to simplify the notation, we will
omit to indicate this dependence in the equations.

At the distances probed by the visible and near-IR coronagraphic images,
the thermal emission from the grains contributes little to the disk brightness
and in the following we will only consider the single scattering of the
stellar light by the dust particles.
$\Phi(r,\phi)$ reads then $\Phi^*\sigma\dma{sca}r^{-2}\,f(\phi)$ where
$\Phi^*$ is stellar flux at Earth at the considered wavelength,
$\sigma\dma{sca}$ and $f(\phi)$ are respectively the scattering cross
section and the scattering phase function of the grains averaged
over the grain size distribution.  Here we implicitly assumed
that the line supporting any incident stellar ray is
an axis of symmetry for the light scattered by the grains (phase function
depending only on $\phi$). This is achieved for instance
when the grains are spherical and homogeneous.

We assume the disk is symmetric about the midplane and the $y$-axis.
Equation~(\ref{eq1}) then re-writes
\begin{eqnarray}
B(y) & = &\Phi^*\int_{0}^{+\infty} \sigma\dma{sca}\,\rho(r',\theta',z')
\,\frac{f(\phi)+f(\pi-\phi)}{r^2} \d x  \\
 & = &\Phi^*\int_{y}^{+\infty} \sigma\dma{sca}\,\rho(r',\theta',z')
\,\frac{f(\phi)+f(\pi-\phi)}{r^2} \frac{\partial x}{\partial r'} \, \d r' \, .
\label{eq4}
\end{eqnarray}
The infinite bound on $r'$ in Eq.\,(\ref{eq4}) can be replaced with the maximal
disk radius $r'\dma{max}$ since the dust density is assumed
to be null beyond $r'\dma{max}$. One can show that 
$\partial x/\partial r' = r'\cos^{-2}i/\sqrt{r^2-y^2}$ 
for $0\leq\phi\leq\pi/2$ with
$\sqrt{r^2-y^2} = \sqrt{r'^2-y^2}/\cos i$.
We moreover have $r^2=\left(r'^2-y^2\sin^2 i\right)/\cos^2i$
when $z=0$.
The scattered light surface brightness
of the disk at a distance $y$ from the star along the $y$-axis thus reads
\begin{eqnarray}
B(y) & = & \Phi^*\,\int^{r'\dma{max}}_{y}
\sigma\dma{sca}\,\rho(r',\theta',z') \nonumber \\
& & \times \frac{f(\phi)+f(\pi-\phi)}{r'^2-y^2\sin^2i}
\,\frac{\cos i}{\sqrt{r'^2-y^2}} \, r'\d r' \, .
\label{Abel}
\end{eqnarray}
In the previous equation, we left the scattering cross section under
the integral sign since the dust properties (composition and/or size distribution)
could, in principle, depend on the distance from the star.  The grain size
distribution in the \bp\ disk is for instance expected to be distance-dependent
\citep[Fig. 4 in][]{aug01}
and the recent observations by \citet{gol06} support this model.
The situation might be similar
in the case of \mic\ as suggested by the observed gradual change of disk
color in scattered light along the disk midplane
\citep[][see also Sect.~\ref{grainsize}]{kri05,met05}.
%
%----------------------------------------------------------------------------
\subsection{The case of edge-on geometrically thin disks}
\label{edgeon}
The \mic\ disk is seen edge-on which simplifies  Eq.\,(\ref{Abel})
($i$=$0$, $r$=$r'$, $\theta$=$\theta'$=$-\phi$ and $z$=$z'$=$0$).
The brightness profiles shown in Fig.\,\ref{Liu_profile} have
been obtained by integrating the disk surface brightness over a few
pixels in direction orthogonal to the disk midplane.
Therefore, $B(y)$ as defined by Eq.~(\ref{Abel}) does not directly
compare with the measured profile and the integration over $z$ has to be
taken into account.

We can take advantage of the integration over
$z$ to get an estimate of the surface number density of the disk instead
of the number density in the disk midplane.
The grains that contribute to $S(y)$ in geometrically thin disks
have their height $z$ above the disk midplane that obeys
$z^2\ll r^2$ which implies $\phi\simeq-\theta$.
Because of the symmetries of the phase function we in addition
have  $f(-\theta)=f(\theta)$ and $f(\pi+\theta)=f(\pi-\theta)$
[see also Eq.\,(\ref{phase})]. 
We approximate the scattered light surface brightness profile measured
along the disk midplane of an edge-on disk and integrated over
$\Delta z/2 \ll r$ in a direction perpendicular to the midplane 
by
\begin{eqnarray}
S(y)
 \simeq \Phi^*\,\int^{r\dma{max}}_{y}
(f(\theta)+f(\pi-\theta))\frac{\sigma\dma{sca}\Sigma(r,\theta)}{r\sqrt{r^2-y^2}
D_z(r,\theta)}
\,\, \d r
\label{Abel3}
\end{eqnarray}
\begin{eqnarray}
\mathrm{with \,\,} & \Sigma(r,\theta) & = \int_{-\infty}^{+\infty}\rho(r,\theta,z)\d z \\
\mathrm{and \,\,} & D_z(r,\theta) & = \Delta z \,\frac{\Sigma(r,\theta)}{\int_{-\Delta z/2}^{+\Delta z/2} \rho(r,\theta,z)\,\,\d z}
\end{eqnarray}
where $\Sigma(r,\theta)$ is the disk surface number density.
Equation\,\ref{Abel3} makes the link between the observed brightness profile
$S(y)$ and the disk properties (surface number density,
phase function and scattering cross section). It depends through
$D_z(r,\theta)$ on the vertical height above the disk midplane over
which the brightness profile has been integrated. Equation\,\ref{Abel3} belongs to
the class of first-kind Volterra integral equations but because of the singularity
in $r=y$, it actually belongs to the sub-class of Abel integral equations.
%
%----------------------------------------------------------------------
\section{Inversion of the \mic\ brightness profiles}
\label{method}
To infer the surface density profile of the \mic\ disk from observed
scattered light profiles, we adopt a direct approach that consists
in performing a basic numerical inversion of Eq.\,(\ref{Abel3}).
This presents several advantages over a classical fitting approach.
In particular, this limits the number of free parameters since the
inversion does not depend on any parametrization of the surface density
profiles. We can therefore efficiently explore the coupling between
the surface density and the anisotropic scattering properties of the grains.
\subsection{Observed surface brightness profiles}
\label{Hband}
\begin{figure}
\centering
\hspace*{-0.3cm}
\includegraphics[angle=0,origin=bl,width=1.04\columnwidth]{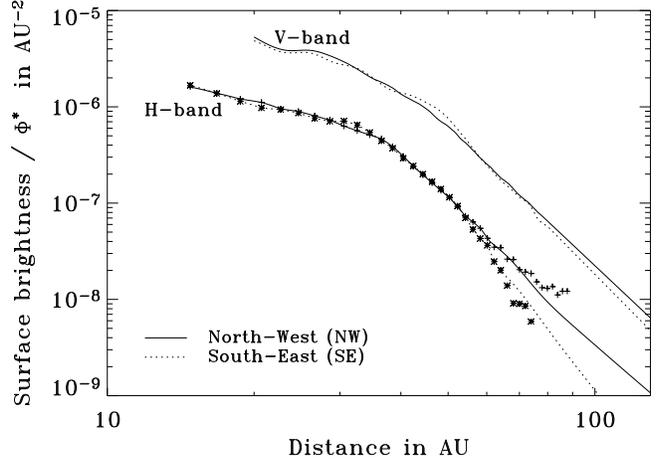}
\caption{Scattered light surface brightness profiles of the \mic\ disk $S(y)$
(divided by the stellar flux $\Phi^*$ to allow comparison) in
the H-band \citep{liu04a} and in the ACS V-band \citep{kri05}.
Beyond $r=65\,$AU, the original H-band profiles have been extrapolated with
radial power laws (NW: $r^{-4.3}$, SE: $r^{-6.1}$).
The visible profiles have been similarly extrapolated beyond 75\,AU
(NW: $r^{-4.7}$, SE: $r^{-5.1}$). The vertical shift between the
two profiles has two origins: the disk color in scattered light
(Sect.\,\ref{grainsize}), and the difference in spatial extent ($\Delta z$)
in a direction orthogonal to the disk midplane used to measure the
disk brightness.
}
\label{Liu_profile}
\end{figure}

In this study we consider two sets of scattered light brightness
profiles ($S(y)$), both displayed in Fig.~\ref{Liu_profile}. The
first set of profiles has been measured by \citet{liu04a}
in the H--band ($\lambda\dma{c} = 1.63\,\mu$m) at the Keck II 10-m
telescope equipped with an adaptive optics system. A spatial resolution
of 0.04\arcsec, or 0.4\,AU at the distance of \mic\ ($9.94\pm 0.13$\,pc),
has been achieved and the profiles have been derived from an aperture
$\Delta z=0.6\arcsec$ ($\sim6$\,AU) wide in the direction perpendicular
to the disk midplane. The Keck H--band coronagraphic images probe the disk
down to $\sim 15$\,AU.
The second set of scattered light brightness profiles has been
obtained in the visible with the HST/ACS instrument by \citet{kri05}.
The ACS images have a resolution of $\sim0.63$\,AU and have been obtained with
$\Delta z=0.25\arcsec$ ($\sim2.5\,$AU). The HST/ACS images present several
features at distances closer than about $20\,$AU that are suspect according
to \citet{kri05}. The published ACS surface brightness profiles were multiplied
by a factor of $1.125$ to account for some confusion regarding the relation to the
ACS geometric distortion and the photometric calibration points assumed by
the HST Synphot package (Krist 2006, private communication).

The HST/ACS visible and Keck-AO near-IR images of the \mic\ disk
show some differences discussed in \citet{kri05}. Their impact on the
surface density profiles calculated with our inversion technique are
addressed in Sect.\,\ref{aumic}.
All observed brightness profiles were oversampled 
to reach a common resolution ($h$ in Appendix \ref{algo}) of $0.2$\,AU.
The observed profiles were furthermore extrapolated with radial power
laws beyond 65\,AU for the H-band profiles and beyond 75\,AU for the visible
profiles (Fig.~\ref{Liu_profile}).
The profiles in Fig.\,\ref{Liu_profile} have been obtained assuming
$\Phi^* = 12.7\,$Jy in the H-band and $\Phi^* = 1.27\,$Jy in the
HST/ACS V-band (F606W).
%
%---------------------------------------------------------------------
\subsection{Scattering phase function}
The scattering phase function of solid particles can be very complex depending
on their shape and composition.
In order to efficiently explore the impact of the anisotropic
scattering properties of the grains [term $f(\theta)+f(\pi-\theta)$
in Eq.~(\ref{Abel3})] on the surface density profiles calculated with
the inversion procedure, we adopt the single parameter
\citet{hen41} phase function 
\begin{eqnarray}
f(\phi) & = & \frac{1-g^2}{4\pi \left(1+g^2-2g\cos\phi \right)^{3/2}}
\label{phase}
\end{eqnarray}
where the parameter $g=\iint_{4\pi} f(\phi)\cos{\phi} \, \d\Omega $ is referred to as
the mean asymmetry factor ($g=0$
for isotropic scattering and $|g|=1$ for purely forward/backward scattering).
The $g$ parameter is related to the scattering properties of the
individual grains through the following relation that ensures that the total
scattered flux is preserved
\begin{eqnarray} 
g  =  \int_{s\dma{min}}^{s\dma{max}} \pi s^2 Q\dma{sca} \times \mathpzc{g}  \, \d n(s)\, / \, \sigma\dma{sca} \label{meang} \\
\mathrm{with\,\,\,\,\,}
\sigma\dma{sca}  =  \int_{s\dma{min}}^{s\dma{max}} \pi s^2 Q\dma{sca}  \d n(s)
\label{sigmasca}
\end{eqnarray}
where $s$ is the grain size, $s\dma{min}$ and $s\dma{max}$ are the minimum
and maximum grain sizes, respectively, and $\d n(s)$ is the unity normalized
differential grain size distribution ($\int_{s\dma{min}}^{s\dma{max}}\d n(s)=1$).
$Q\dma{sca}$ is the dimensionless scattering
efficiency and $\mathpzc{g}$ is the asymmetry factor that depends on the grain
properties (size, composition, structure, ...) and on the wavelength.

To efficiently explore the impact of scattering anisotropy on the results
and to limit the number of free parameters, we will assume that the $g$
parameter is a constant.
The inversion algorithm can nevertheless account for mean asymmetry parameters
that depend on the grain position in  the disk.
We will also only consider the absolute value of $g$
since we cannot distinguish between positive and negative values
due to the edge-on orientation.
The asymmetry factors $|g\uma{H}|$ and $|g\uma{V}|$ in the H-band
and ACS-V-band, respectively, are hence the only free
physical parameters when reconstructing surface density
profiles from scattered light profiles using Eq.\,(\ref{Abel3}).
\subsection{Vertical and azimuthal profiles}
\label{vprofile}
Following \citet{kri05}, we assume that the vertical profile of the
disk density can be approximated by a Lorentzian function
$\left(1+z^2/H^2(r)\right)^{-1}$. The scale height $H(r)$ is parametrized
as follows: $H(20\,{\rm AU}) = 0.8675$\,AU, $H(r)\propto r^{0.08}$
for $r\leq 49\,$AU, $H(r)\propto r^{2.5}$ for $49< r \leq 80\,$AU 
and $H(r)\propto r^{0.3}$ for $r>80\,$AU.
$D_z(r,\theta)$ thus writes
\begin{eqnarray}
D_z(r,\theta) = \Delta z\,\frac{\pi}{2}\arctan^{-1}{\left(\frac{\Delta z}{2\,H(r)}\right)}
\end{eqnarray}
such that $D_z(r,\theta) = \Delta z$ for $\Delta z \gg 2\,H(r)$.
\begin{figure*}[tbp]
\centering
\hbox to \textwidth
{
\parbox{0.5\textwidth}{
\includegraphics[angle=0,width=1.\columnwidth,origin=bl]{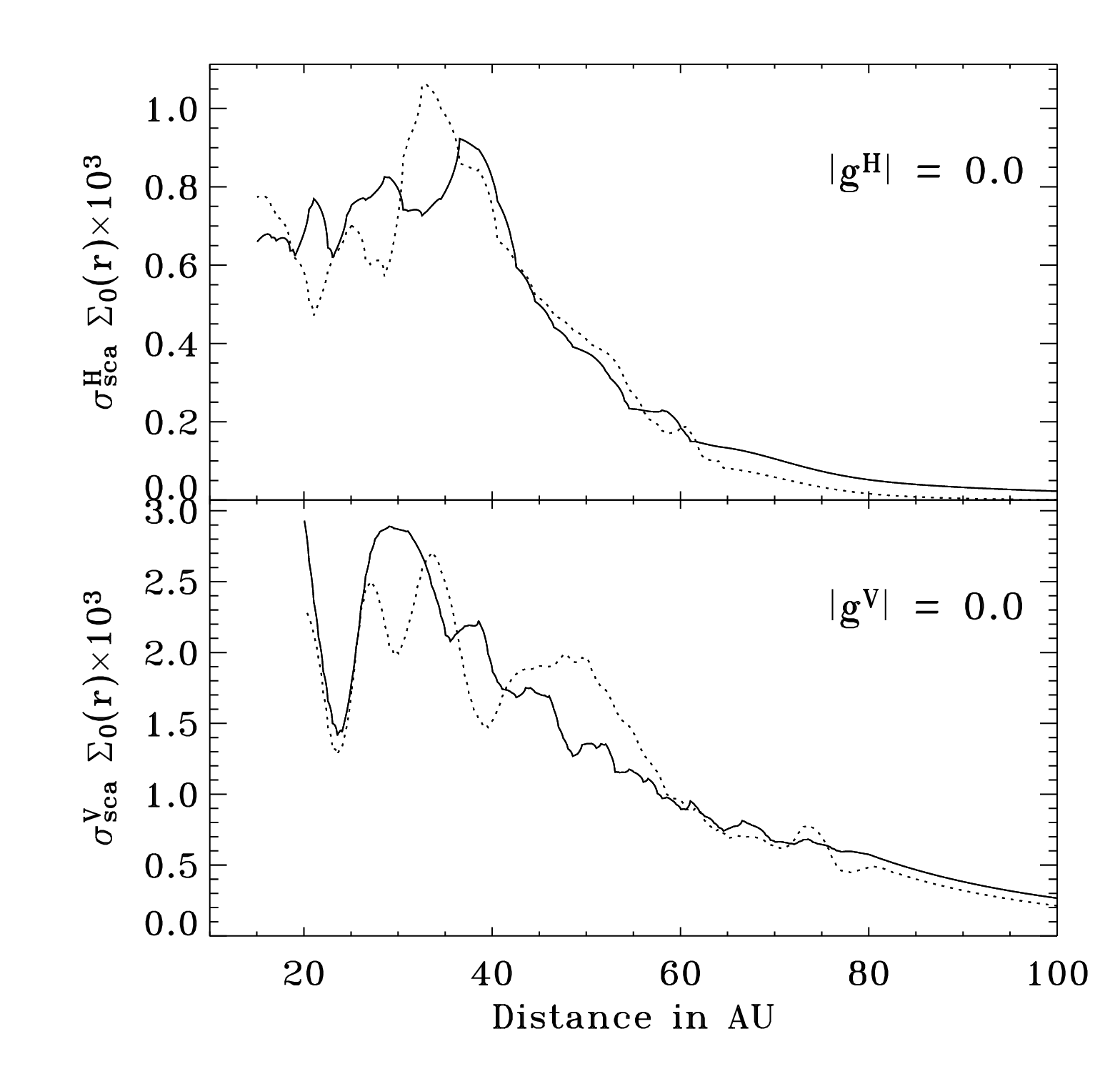}
}
\parbox{0.5\textwidth}{
\includegraphics[angle=0,width=1.\columnwidth,origin=bl]{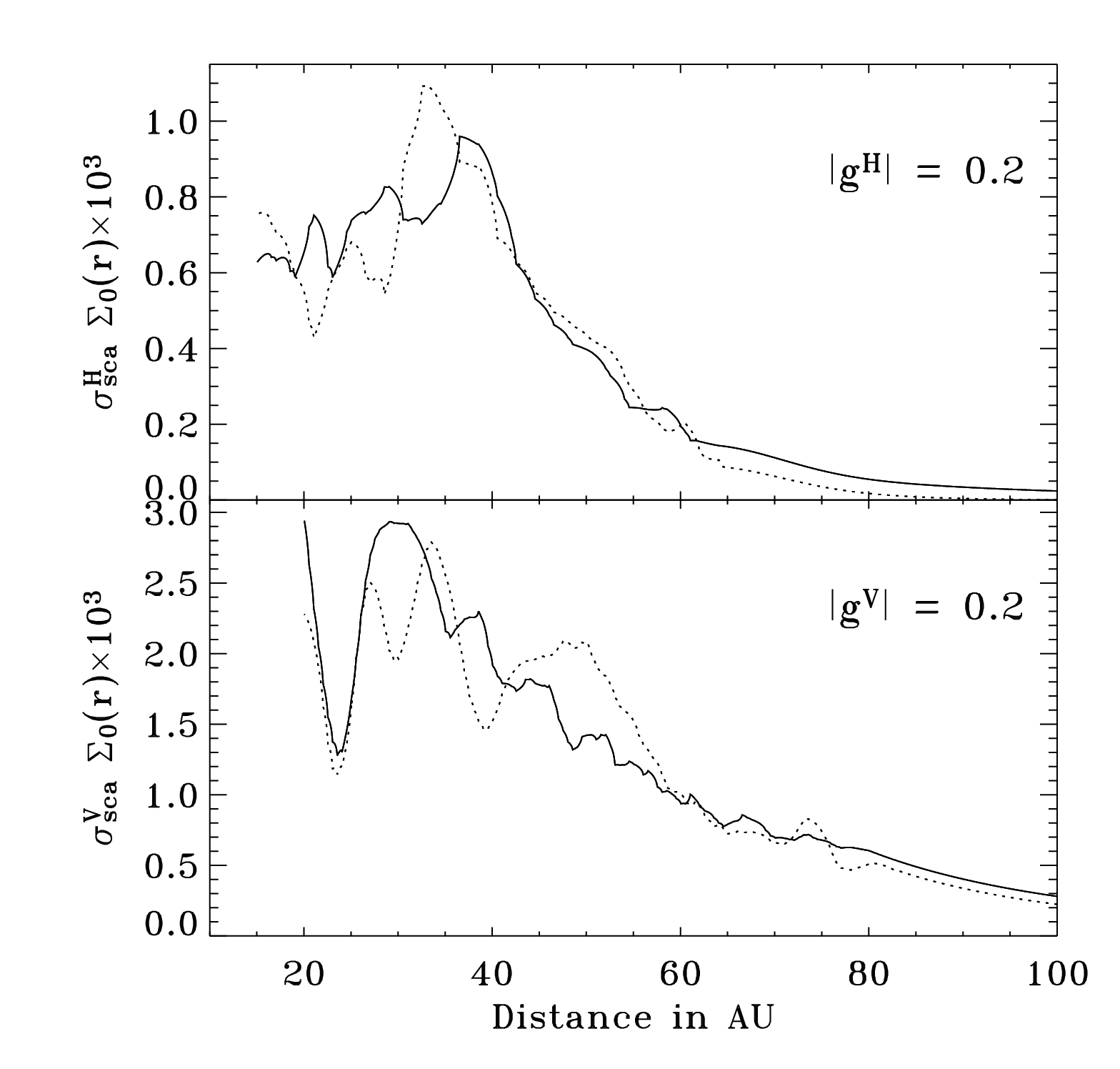}
}
}
\hbox to \textwidth
{
\parbox{0.5\textwidth}{
\includegraphics[angle=0,width=1.\columnwidth,origin=bl]{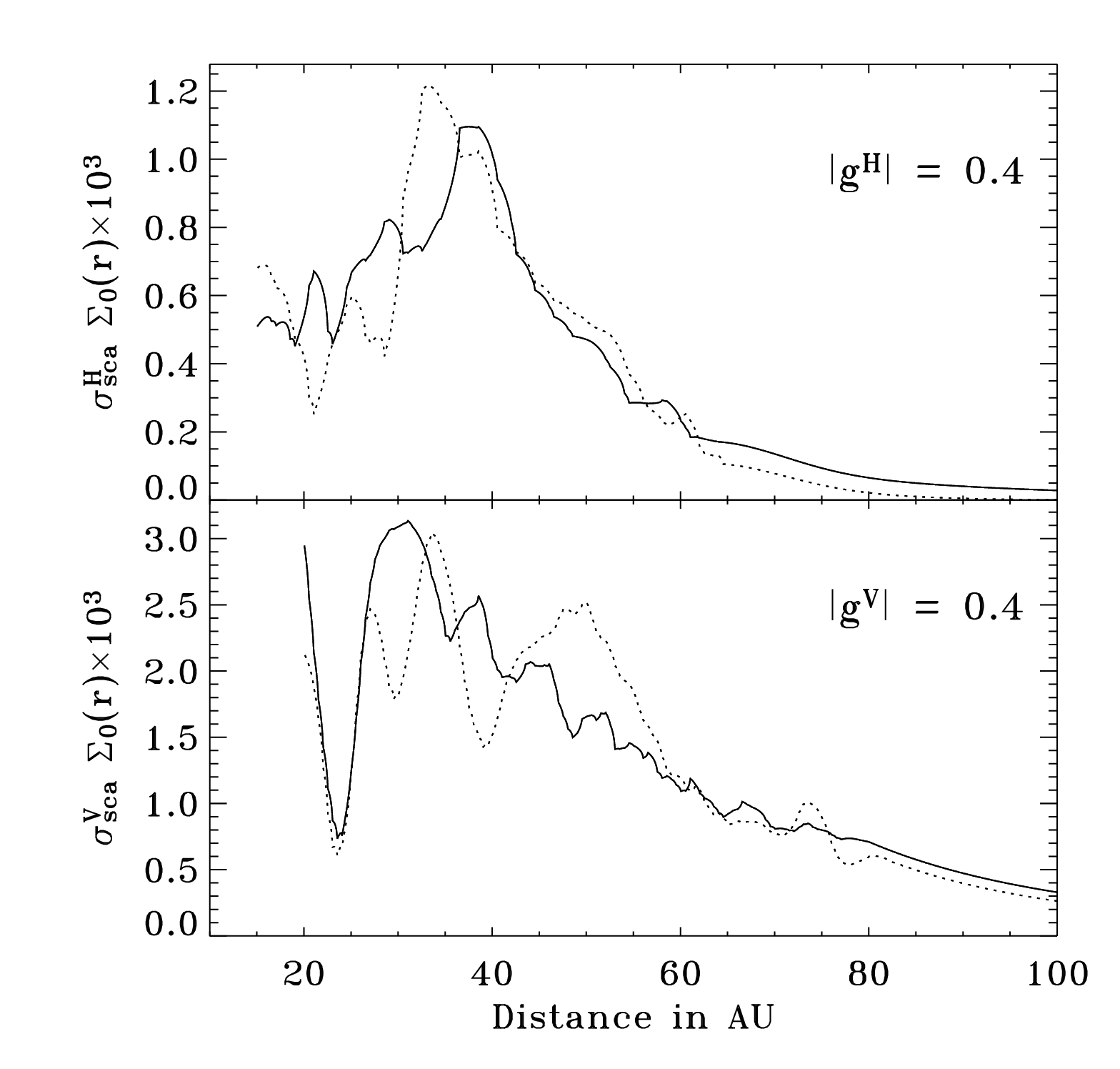}
}
\parbox{0.5\textwidth}{
\includegraphics[angle=0,width=1.\columnwidth,origin=bl]{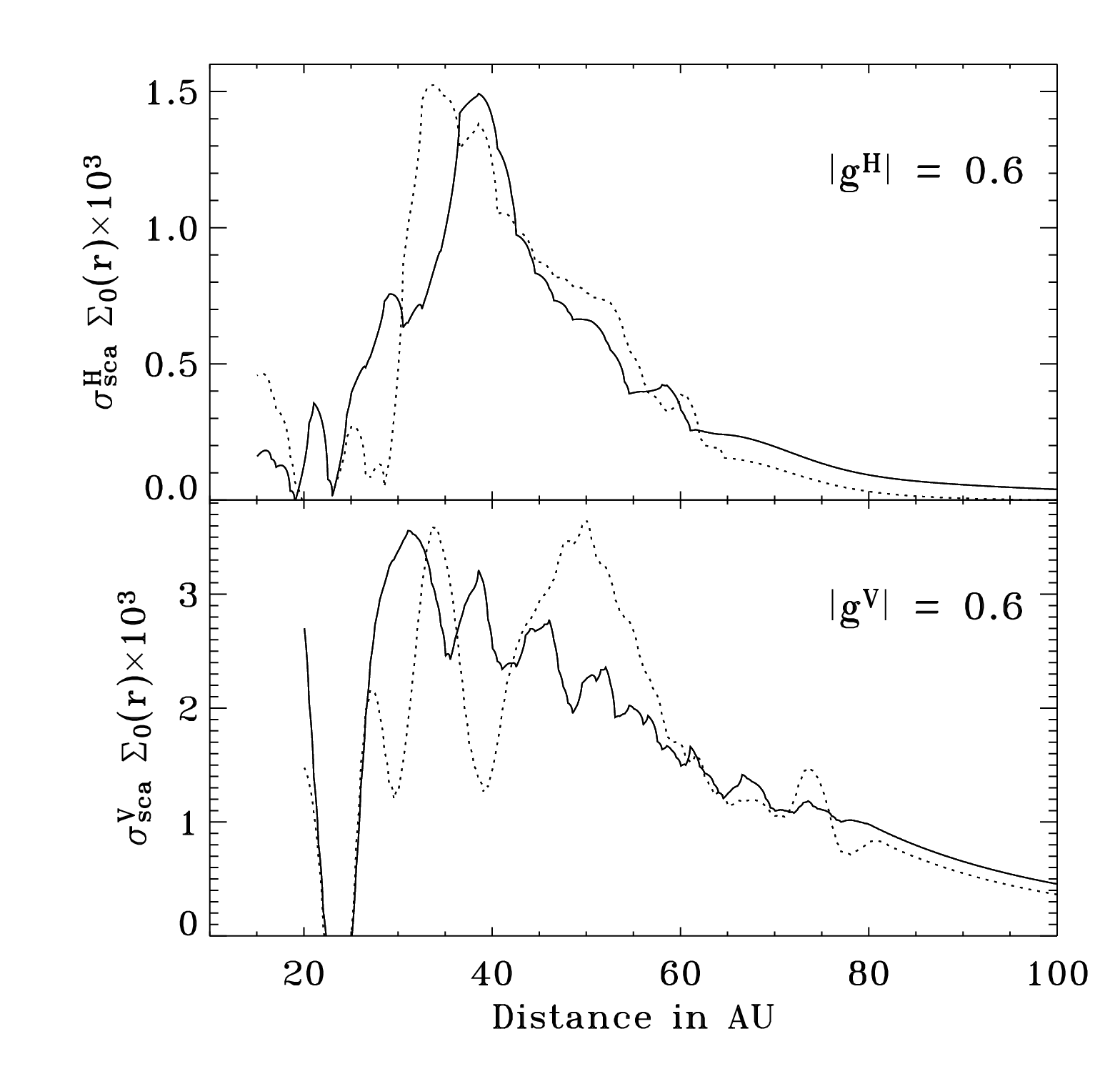}
}
}
\caption{Surface density profiles obtained by inverting the \citet{liu04a}
H-band observations and the \citet{kri05} HST/ACS visible observations of the
\mic\ disk (top and bottom panels, respectively, for each figure), assuming 
four mean scattering asymmetry parameters. The solid line corresponds
to the NW side; the dotted line to the SE side.
These results have been obtained assuming $\delta=1$ (see Sect.\,\ref{vprofile}
and App.\,\ref{theta}) but the profiles are
poorly sensitive to the assumed azimuthal profile provided $\delta \lesssim 5$.
}
\label{midplane}
\end{figure*}

The scattered light images of the \mic\ disk evidence side-to-side
asymmetries.
Also, to appreciate how the inversion of the NW brightness profile
affects the inversion of the SE brightness profile (and reciprocally),
we introduce an {\it ad hoc} dependence on $\theta$ in the inversion
process and we proceed to the inversion of the NW and SE brightness
profiles simultaneously. We re-write the dust
surface number density as
$\Sigma(r,\theta)= \Sigma_0(r)\Theta(r,\theta)$
where $\Theta(r,\theta)$ is a unit-less function given in Appendix \ref{theta}
and characterized by a single, purely geometrical, parameter $\delta$.
The role of the $\Theta(r,\theta)$ function is to smoothly
link the NW and SE sides of the disk. During the inversion process,
the surface density is calculated approximately along the $y$--axis
(see Fig.~\ref{sketch}).
The  $\Theta(r,\theta)$ function forces the NW and SE surface density
profiles to be identical along the $x$--axis and equal to their mean value.
%
%
%----------------------------------------------------------------------------
\subsection{Surface density profiles}
%----------------------------------------------------------------------------
\label{aumic}

To reconstruct the surface density profiles of the disk,
we have used a classical numerical scheme to invert Eq.\,(\ref{Abel3}).
The algorithm is detailed in Appendix~\ref{algo} in the
case of edge-on disks but it can easily be extended to the case
of inclined disks \citep[e.g. the ring about the F5 star
\hd181, ][]{sch06}.
As a sanity check, the calculated surface density profiles were
systematically incorporated into our 3D debris disk model
to compute synthetic brightness profiles \citep{aug99,aug01}. These
synthetic profiles were always found to be in excellent agreement
with the original observed profiles.

The results of our inversion procedure are reported in Figure~\ref{midplane}.
This figure displays the various surface density profiles found
assuming four different values for the mean asymmetry parameter $|g|$.
For each $|g|$ value, the top panel shows the surface density profile that
reproduces the H-band scattered light surface brightness profile measured
by \citet{liu04a}, while the profile in the bottom panel is derived from
the ACS-V-band observations \citep{kri05}.
The surface density profiles are found to depend quite sensibly
on the assumed $|g|$ value. Large $|g|$ values promote the
emission at scattering angles close to $0$ or $\pi$ (depending on the
sign of $g$) which is achieved when the grains are at distances
from the star large compared to their projected distance
($r\gg y$ in Eq.\,(\ref{Abel3})).
The amount of dust at large distances thus tends to increase
with increasing $|g|$ values and the regions interior to the
disk thus become more and more dust-depleted. As a result,
the disk as derived from the H-band brightness profiles resembles
a ring peaked around $35$\,AU for sufficiently large $|g\uma{H}|$ values.
Furthermore, $|g|$ cannot be significantly
larger than $\sim 0.6$ since for larger $|g|$ values, the inversion
algorithm requires the surface density profiles to be negative at some
radii to compensate for the too large flux, compared to the observations,
intercepted on the line of sight.

There are obvious differences between the surface density profiles obtained from
the two data sets used in this study. They basically reflect the differences
between the original images \citep[see discussion in][]{kri05}. Along a given
arm, the same density features can generally be identified
(though often shifted by $1$--$2$\,AU with respect to each other) but not
with the same strength. For instance, the NW surface density profiles inferred
from the H-band observations, peak at $\sim 37\,$AU and show a smaller peak
around $30$\,AU, while the feature strengths are reverted in the case of the
surface density profiles obtained from the visible images. These two peaks
correspond to features (D) and (C), respectively, according to the nomenclature
of \citet{liu04a},
and feature (D) is only marginally resolved in the ACS images. Along the SE
side, features (A) and (B) around $25$--$27$\,AU and $33$--$34$\,AU, respectively,
show strength ratios that depend on $|g|$ and on the data set used to calculate
the disk surface density. The visible images in addition harbor an extended
clump around $50$\,AU that results in a broad feature at that distance,
and which is only marginally present in the surface density profiles deduced
from the H-band observations. Almost all profiles, nevertheless, show
a common dust depletion around $22$--$24\,$AU, the exact position
depending on the data set used to calculate the surface density profiles.
Besides the differences, we note that all profiles show complex structures
that can hardly be approximated with simple power laws. 

\begin{figure}
\centering
\hspace*{-0.7cm}
\includegraphics[angle=0,origin=bl,width=1.1\columnwidth]{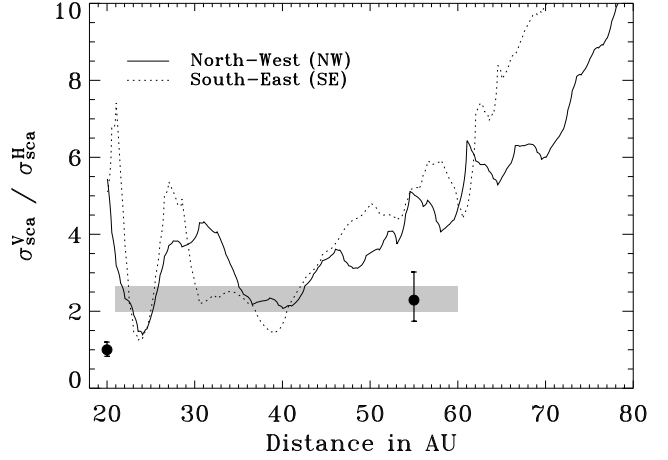}
\caption{Scattering cross section ratio $\sigma\dma{sca}\uma{V} /
\sigma\dma{sca}\uma{H}$ obtained by dividing the profiles calculated
for $|g\uma{H}| = |g\uma{V}| = 0.4$  and shown in Fig.\,\ref{midplane}.
The gray area corresponds to the range of scattering cross section ratios
inferred from the analysis of the disk color in the ACS bands
(Sect.\,\ref{grainsize} and Table\,\ref{nominal}). The R$-$H colors inferred
by \citet{met05} are displayed as two black dots at $20$ and $55$\,AU.
}
\label{scaratio}
\end{figure}

The differences between the surface density profiles derived from the visible
and infrared observations may result from a combination of instrumental effects
and grains optical properties varying throughout the disk midplane.
The inversion of additional observed profiles and more constraints on
the grain properties would help to disentangle
between these two possibilities.
But if we assume that the differences are real, then the ratio
$\sigma\dma{sca}\uma{V}\Sigma_0(r) / \sigma\dma{sca}\uma{H}\Sigma_0(r)$ is
an estimate of the V- to H-band scattering cross section ratio as a function
of the distance. This ratio is plotted in Figure\,\ref{scaratio} in the
case $|g\uma{H}| = |g\uma{V}| = 0.4$. This suggests that the
$\sigma\dma{sca}\uma{V} / \sigma\dma{sca}\uma{H}$ ratio gradually increases
beyond $\sim 40\,$AU, in line with the color gradient of the disk over the same
region \citep{kri05}. Qualitatively,
this would indicate that the large grains become less and less abundant,
relative to the smallest grains at the same position, with increasing the
distance from the star.
This behaviour is expected if the grains have not been produced
locally but placed in orbits of high eccentricity
by a pressure force $\propto s^2/(r^2+z^2)$ (where $s$ is the grain size)
as shown in Fig.~4 of \citet{aug01}.
We discuss in Sect.\,\ref{rad} and \ref{wind} the origin of pressure
forces in the case of \mic.
\begin{figure*}[tbp]
\centering
\hbox to \textwidth
{
\parbox{0.5\textwidth}{
\includegraphics[angle=0,origin=bl,width=\columnwidth]{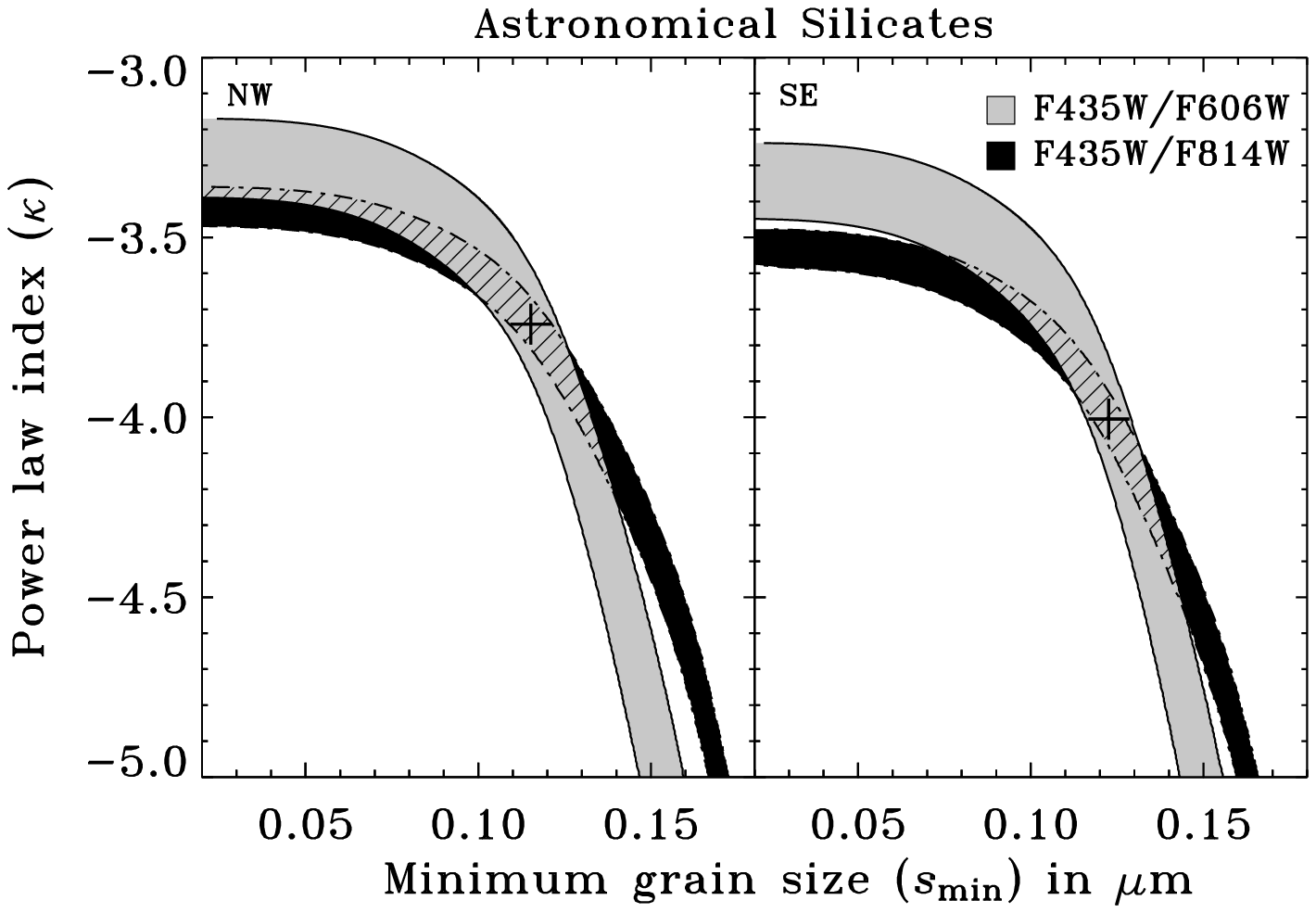}
}
\parbox{0.5\textwidth}{
\includegraphics[angle=0,origin=bl,width=\columnwidth]{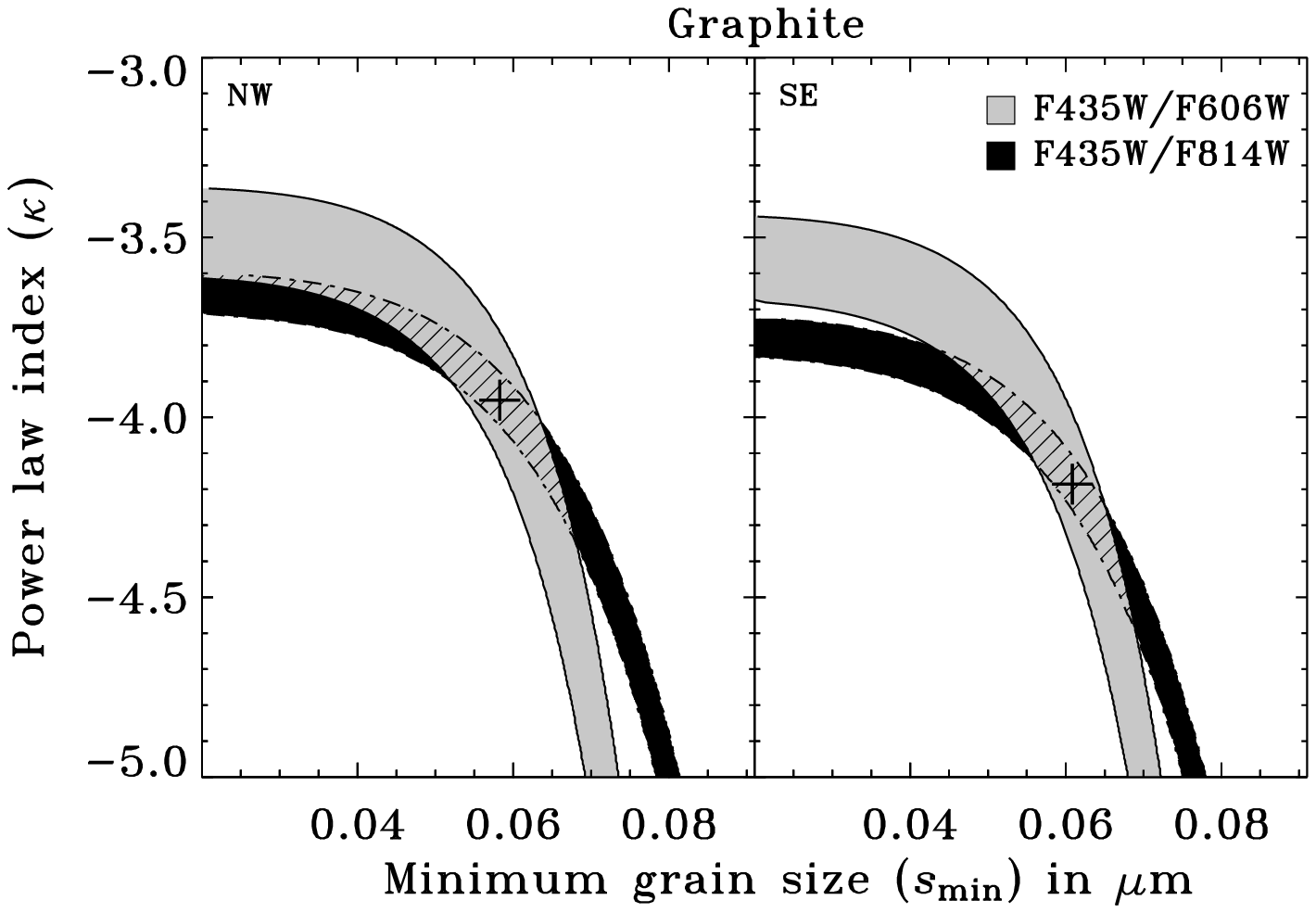}
}}
\caption{Minimum grain sizes $s\dma{min}$ and power-law indexes $\kappa$
of the size distribution consistent with the HST/ACS colors measured
by \citet{kri05}.
Gray area: F435W/F606W, dark area: F435W/F814W, dashed area: intersection.
The cross sign corresponds to the nominal pair ($s\dma{min}$, $\kappa$) 
that produces a perfect match to the two colors.
The nominal values are given in Table~\ref{nominal}.}
\label{col}
\end{figure*}
\begin{table*}[btp]
\begin{center}
\caption{Grain properties inferred from the interpretation of the disk colors
in the visible \citep{kri05} and implications for the disk mass.}
\label{nominal}
\begin{tabular}{llcccccl}
\hline
\hline
\noalign{\smallskip}
 &  & \multicolumn{2}{c}{\sc Astronomical Silicates} & & \multicolumn{2}{c}{\sc Graphite Grains} & 
 \\ 
\noalign{\smallskip}
\cline{3-4} \cline{6-7}
\noalign{\smallskip}
\noalign{\smallskip}
Description & Parameter$^{(\mathrm{a})}$  & NW & SE & & NW & SE & [unit]\\
\noalign{\smallskip}
\hline
\noalign{\smallskip}
Nominal minimum grain size & $s\dma{min}$ &
$0.115$ &
$0.123$ &
& 
$0.0583$ &
$0.0609$ &
[$\mu$m] \\
Nominal power law index 
& $\kappa$  
&
$-3.74$ &
$-4.01$ &
&
$-3.95$ &
$-4.18$ &
\\
H-band mean asymmetry factor & $|g\uma{H}|$  &
$0.637$ &
$0.604$ &
&
$0.358$ &
$0.318$ &
\\
V-band mean asymmetry factor & $|g\uma{V}|$  &
$0.675$ &
$0.661$ &
&
$0.455$ &
$0.421$ &
\\
H-band mean scattering cross section & $\sigma\dma{sca}\uma{H}$ &
$16.0$  & 
$12.7$ & 
&
$2.12$ & 
$1.64$ & 
[$10^{-10}$\,cm$^2$] \\
Ratio of mean cross sections & $\sigma\dma{sca}\uma{V} / \sigma\dma{sca}\uma{H}$  &
$2.05$ & 
$2.56$ & 
&
$2.15$ & 
$2.65$ & 
 \\
V-band mean albedo & $\omega\uma{V}$ &
$0.795$  & 
$0.822$ &
&
$0.518$ &
$0.513$ &
 \\
\hline
\noalign{\smallskip}
\hspace*{0.5cm} &  & \multicolumn{5}{c}{\sc H-band data set} & \\ %
\noalign{\smallskip}
\cline{3-7}
\noalign{\smallskip}
Disk scattering cross section$^{(\mathrm{b})}$  & $S\dma{sca}\uma{H}$ &
$4.93 - 6.97 $ & 
$4.51 - 6.30 $ &
&
$4.93 - 6.97 $ &
$4.51 - 6.30 $ &
[AU$^2$] \\
Dust mass$^{(\mathrm{c})}$ &  &
$2.60 - 3.67$ &
$0.952 - 1.33$ &
&
$0.583 - 0.824$ &
$0.306 - 0.428$ &
[$10^{-4} M_{\oplus}$] \\
\noalign{\smallskip}
\cline{3-4} \cline{6-7}
\noalign{\smallskip}
{\sc Total dust disk Mass}$^{(\mathrm{c})}$ & $M\dma{disk}$ & \multicolumn{2}{c}{$3.55 - 5.00$} & & \multicolumn{2}{c}{$0.889 - 1.25$} & [$10^{-4} M_{\oplus}$]
 \\ 
\hline
\noalign{\smallskip}
\hspace*{0.5cm} &  & \multicolumn{5}{c}{\sc V-band data set} & \\ %
\noalign{\smallskip}
\cline{3-7}
\noalign{\smallskip}
Disk scattering cross section$^{(\mathrm{b})}$  & $S\dma{sca}\uma{V}$ &
$15.9 - 23.7 $ & 
$15.5 - 23.1 $ &
&
$15.9 - 23.7 $ & 
$15.5 - 23.1 $ &
[AU$^2$] \\
Dust mass$^{(\mathrm{c})}$ &  &
$4.10 - 6.10$ &
$1.28 - 1.91$ &
&
$0.876 - 1.31$ &
$0.398 - 0.593$ &
[$10^{-4} M_{\oplus}$] \\
\noalign{\smallskip}
\cline{3-4} \cline{6-7}
\noalign{\smallskip}
{\sc Total dust disk Mass}$^{(\mathrm{c})}$ & $M\dma{disk}$ & \multicolumn{2}{c}{$5.38 - 8.01$} & & \multicolumn{2}{c}{$1.27 - 1.90$} & [$10^{-4} M_{\oplus}$]
 \\ 
\hline
\end{tabular}
\end{center}
\begin{list}{}{}
\item[$^{\mathrm{a}}$]  H and V superscripts 
stand for H--band and ACS V--band, respectively
\item[$^{\mathrm{b}}$] for $0\leq |g| \leq 0.6$ and $\delta = 1$. $S\dma{sca}$,
given by Eq.~(\ref{Ssca}), does not depend on the assumed grain composition
\item[$^{\mathrm{c}}$] mass inferred from the analysis of the scattered light
observations. See Sect.\,\ref{mass} for an estimate of the dust mass
deduced from the analysis of the disk thermal emission. Mass obtained
assuming $s\dma{max}=1$\,mm, $\rho\dma{Silicates} = 3.5$\,g.cm$^{-3}$
and $\rho\dma{graphite} = 2.2$\,g.cm$^{-3}$.
The total dust disk mass is the sum of the NW and SE dust masses.
\end{list}
\end{table*}
%
%----------------------------------------------------------------------------
\section{Grain and Disk Properties}
%----------------------------------------------------------------------------
%
\label{diskprop}
\subsection{Disk color and grain size distribution}
\label{grainsize}

The output of the inversion procedure is not the actual surface density
profile of the disk, $\Sigma_0(r)$, but its product by the mean scattering cross
section $\sigma\dma{sca}$. To evaluate $\sigma\dma{sca}$, and hence $\Sigma_0(r)$,
we search for the grain properties that can reproduce the blue color of
the \mic\ disk reported by \citet{kri05} in the HST/ACS bands. These
authors find F435W/F606W
flux ratios, relative to the star and averaged over a $43\,$AU long
aperture centered at $r=42\,$AU, of $1.13\pm 3\%$ and $1.15\pm 3\%$
for the NW and the SE sides of the disk, respectively. The  measured F435W/F814W
flux ratios amounts to $1.35\pm 3\%$ (NW) and $1.44\pm 3\%$ (SE).
Qualitatively, the blue color indicates that the visible
scattered light images are dominated by elementary (sub-)grains
of the order or smaller then a few tenths of micrometers.

Provided the phase functions are not too different,
the measured colors are only dependent upon the ratios of
mean scattering cross sections in the ACS passbands.
We have considered optical constants for astronomical silicates \citep{wei01}
and graphite grains \citep{lao93}, and we used the Mie theory,
appropriate for hard spheres, to calculate the scattering efficiency $Q\dma{sca}$.
We have assumed a differential grain size
distribution of the form $\d n(s)\propto s^{\kappa}\d s$ with
$\kappa < 0$ between $s\dma{min}$ and $s\dma{max}$. The maximum grain size
is assumed to be large enough to not affect the results. The scattering
cross sections were calculated using Eq.\,(\ref{sigmasca}).

We have searched for the pairs ($s\dma{min}$, $\kappa$) that give
scattering cross section ratios equal to the two HST/ACS colors.
Each side of the disk has been considered independently.
The results are reported in Table~\ref{nominal} and in Fig.~\ref{col}, where
the cross sign indicates the nominal couple ($s\dma{min}$, $\kappa$), namely
the couple that produces a perfect match to the two ACS disk colors.
For astronomical silicates, the blue color of the disk is reproduced
when the minimum grain size $s\dma{min}$ is smaller than $0.14\,\mu$m
with a nominal $s\dma{min}$ value of $0.12\,\mu$m. For graphite grains,
we find $s\dma{min} < 0.07\,\mu$m with a nominal $s\dma{min} \simeq 0.06\,\mu$m.
For a given grain composition, our simple model predicts quite similar nominal
$s\dma{min}$ values for both the SE and NW sides of the disk, but the nominal
slopes  $\kappa$ differ: the size distribution in the NW side is found to be
less steep than SE size distribution where the disk is found to be even bluer
\citep{kri05}. $\kappa$ values consistent with the HST/ACS
colors are generally found to be slightly smaller than the canonical $-3.5$ value
that holds for systems in collisional equilibrium \citep{don69}.
The minimum grain sizes we infer for astronomical silicates are smaller than
the $s\dma{min}$ value of $0.5^{+0.5}_{-0.2}\,\mu$m obtained by \citet{met05}
based on the R$-$H color of the disk measured around 50--60\,AU and assuming
$\kappa = -3.5$.

The mean asymmetry factors reported in Table\,\ref{nominal}
were calculated using Eq.\,(\ref{meang}). For astronomical silicates, $|g|$ reaches
or slightly overcomes the $0.6$ limit derived from the inversion
of the scattered light profiles (Sect.\,\ref{aumic}). For more refractory
grains such as graphite grains, the mean asymmetry factor remains smaller
than the $|g|=0.6$ limit and is close, in the V-band, to the $0.4$ value
inferred by \citet{kri05} from the disk modeling.
\subsection{Disk optical thickness}
\label{tau}
In the previous sections, we have calculated the surface density
and grain properties of the \mic\ disk assuming an optically
thin medium at all wavelengths and in all directions. This
assumption has been justified in Sect.~\ref{hyp} by the 
low fractional disk luminosity and the appearance of the disk
at visible and near-IR wavelengths.
We verify in the following that our results are
self-consistent.

\begin{figure}
\centering
\hspace*{0.1cm}
\includegraphics[angle=0,origin=bl,width=\columnwidth]{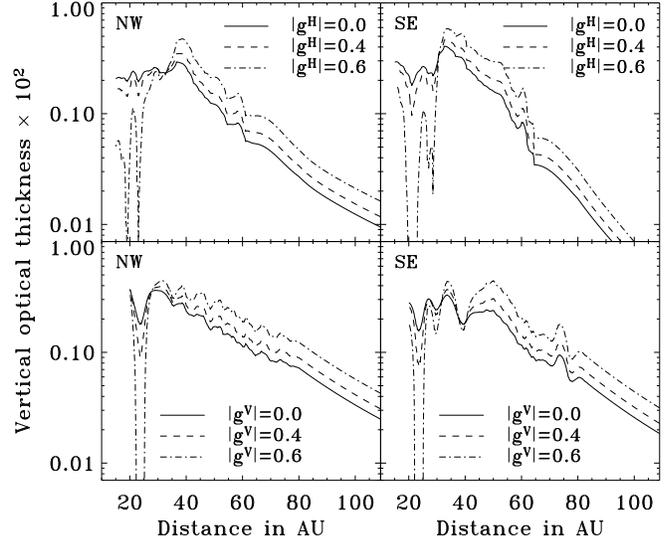}
\caption{Vertical optical thickness assuming
astronomical silicates (see Table~\ref{nominal}).Top panels: surface
density profiles inferred from the inversion of the H-band observations;
bottom panels: surface density profiles derived from V-band observations.}
\label{taufigure}
\end{figure}

The vertical optical thickness of the \mic\ disk, $\tau_{\perp}(r)$, reads
\begin{eqnarray}
\tau_{\perp}(r)  =  \frac{1}{\omega\uma{V}}\times \sigma\dma{sca}\uma{V}\Sigma_0(r) 
= \frac{\sigma\dma{sca}\uma{V}}{\omega\uma{V}\,\sigma\dma{sca}\uma{H}}\,
\times\,\sigma\dma{sca}\uma{H}\Sigma_0(r)
\label{tauper}
\end{eqnarray}
where $\omega\uma{V}$ is the albedo of the dust particles
in the V-band averaged over the grain size distribution.
The ratio $\sigma\dma{sca}\uma{V}/\sigma\dma{sca}\uma{H}$
and the mean albedo $\omega\uma{V}$ have been estimated assuming
the grain properties summarized in Table\,\ref{nominal}.
In the visible, the disk vertical thickness $\tau_{\perp}(r)$
does not exceed $\sim\pdx{5}{-3}$ indicating that the disk
is vertically optically thin (Fig.~\ref{taufigure}).
We checked that with the assumed grain
properties, the vertical thickness is maximum in the visible implying
$\tau_{\perp} \ll 1$ at all wavelengths.

\begin{figure}
\centering
\hspace*{0.1cm}
\includegraphics[angle=0,origin=bl,width=\columnwidth]{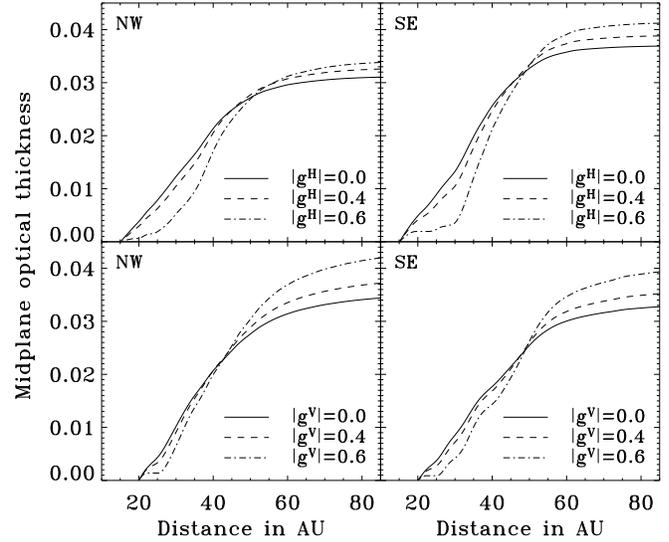}
\caption{Midplane optical thickness assuming
the same grain properties as in Fig.~\ref{taufigure}.}
\label{taumidplanefigure}
\end{figure}

The midplane density of the disk writes $\Sigma_0(r) / (\pi H(r))$ and the
optical thickness along the disk midplane thus reads
\begin{eqnarray}
\tau_{\parallel}(r) = \int_{r\dma{min}}^{r}
\frac{\sigma\dma{sca}\uma{V}}{\omega\uma{V}}\, 
\frac{\Sigma_0(\mathpzc{r})}{\pi H(\mathpzc{r})}\,\d \mathpzc{r} = \int_{r\dma{min}}^{r}
\frac{\tau_{\perp}(\mathpzc{r})}{\pi H(\mathpzc{r})}\,\d \mathpzc{r}
\label{taupar}
\end{eqnarray}
where $r\dma{min}=15\,$AU in the H-band and $r\dma{min}=20\,$AU in the ACS V-band
(see Figs.~\ref{Liu_profile} and \ref{midplane}).
Figure~\ref{taumidplanefigure} shows that
$\tau_{\parallel}(r)$ does not exceed $\sim\pdx{4}{-2}$ in the visible
whatever $|g|\leq 0.6$ and this result is independent of the data set used to
calculate the surface density profiles.  Given that $\tau_{\perp}$ and
thus $\tau_{\parallel}$ both reach their maxima at visible wavelengths in
this model, these results indicate that the disk is optically thin in all
directions and at all wavelengths.
\subsection{Disk mass and implication for the grain properties}
\label{mass}
From the surface density profiles calculated in Sect.~\ref{aumic}
and the constraints on the grain size distribution obtained
in Sect.~\ref{grainsize}, we can estimate the mass of the dust disk
$M\dma{disk}$ further out than $r\dma{min}$
\begin{eqnarray}
M\dma{disk} & = & \frac{S\dma{sca}(|g|)}
{\sigma\dma{sca}}
\times \int_{s\dma{min}}^{s\dma{max}}
\rho\frac{4\pi}{3}s^3\,\d n(s)
\end{eqnarray}
where $\rho$ is the grain mass density. The disk scattering
cross section $S\dma{sca}(|g|)$ writes
\begin{eqnarray}
\label{Ssca}
S\dma{sca}(|g|) & = & \int_{r=r\dma{min}}^{r\dma{max}}
\int_{\theta=0}^{2\pi}\sigma\dma{sca}\Sigma(r,\theta)r\d r\d \theta
\end{eqnarray}
where $r\dma{max}$ was fixed to $140$\,AU.
The results are given in Table~\ref{nominal}.

The total mass of dust derived from the scattered light analysis is of
the order, to a few times, the mass of the Ceres asteroid
($\sim\pdx{1.46}{-4}\,M_{\oplus}$) in the Solar System.
This is about $10$ to $80$ times the estimated dust mass in the
Kuiper Belt \citep[$\sim 10^{-5}\,M_{\oplus}$,][]{bac95}. But it
is significantly smaller than the dust mass calculated by \citet{liu04b}
from sub-millimeter observations ($\pdx{1.1}{-2}\,M_{\oplus}$).
In fact, a dust mass as low as a few $10^{-4}\,M_{\oplus}$ is
not expected at an age of about $10\,$Myr \citep[e.g.][]{naj05}.

To help identifying the origin of the mass difference, we calculated the disk
Spectral Energy Distribution (SED) assuming the surface density profiles and
the grain properties obtained from the analysis of the scattered light
observations. We assumed the star is in a quiescent period (no UV-flares,
see Fig.\,\ref{sed} and Sect.\,\ref{parsed}). We find that the total
$25\,\mu$m flux is always well reproduced while the $60$--$70\,\mu$m fluxes
are systematically underestimated by factors of $1.5$ to $2$. More
critically, the observed sub-millimeter emissions are
$10$ to $20$ times larger than predicted when silicates are considered.
These flux ratios amount to $15$ to $30$ for the graphite grains.

On the other hand, we find that the whole disk SED can be reproduced
provided the constraints on the grain size distribution are relaxed.
Good fits to the SED are for instance obtained for $s\dma{min} \simeq 1.0\,\mu$m
and $\kappa \simeq -3.4$ when silicate grains and the surface density profiles
derived from the H-band images are assumed (Fig.\,\ref{sed}). With the V-band
surface density profiles, we find $s\dma{min} \simeq 0.1\,\mu$m and
$\kappa \simeq -3.4$. In both cases, the dust disk mass is approximately
$\pdx{7}{-3}\,M_{\oplus}$, in better agreement with the submillimeter
dust mass ($\pdx{1.1}{-2}\,M_{\oplus}$)
calculated by \citet{liu04b} assuming a single temperature formula.

The grain size distributions that reproduce the disk colors in scattered
light do then underestimate the amount of large grains in the disk, which
explains the disagreement between the two independent mass estimates. It might
be that the dust properties in Table~\ref{nominal} are only valid
in a small range of grain sizes. This could be the case, for instance, if
the actual grain size distribution departs from a simple power law
as expected for systems in collisional equilibrium
\citep[see discussion in][]{the03}. It might also be possible 
that the grains have complex structures, and hence optical properties,
that cannot be consistently reproduced over the whole spectrum assuming
hard spheres.
%
%
%%%%%%%%%%%%%%%%%%%%%%%%%%%%%%%%%%%%%%%%%%%%%%%%%%%%%%%%%%%%%%%%%%%%%%%%%%
%
%
\section{Diffusion of dust particles in the outer disk.
I.~Time-variable radiation forces}
\label{rad}
Radiation pressure is a key parameter controlling the dynamics
of dust particles in usual debris disks. The grains are
thought to be produced by numerous, colliding or evaporating
planetesimals orbiting the star on nearly circular orbits. 
Depending on their size, the radiation  pressure can transport the grains
far away from their production zone and fill the outer regions
of the disk. This model explains the observed radial profiles
of the \bp\ and \hd181\ disk for instance \citep{lec96,aug01,the05,sch06}.
We discuss below and in the next Section whether
this model could or not apply to the \mic\ disk.
\subsection{Equation of motion and dynamical time-scales}
A dust particle of mass $m$ and geometric cross section $\sigma$
feels a full force from the stellar radiation field of energy flux
$F$ that reads
\begin{eqnarray}
\left.m\,\frac{\d^2\vec{r}}{\d t^2}\right|_{\mathrm{PR}}
=\frac{F\sigma Q\dma{pr}}{c}
\left[\left(1-\frac{\dot{r}}{c}\right)\vec{\hat{u}}-\frac{\vec{v}}{c}\right]
\label{fullrp}
\end{eqnarray}
\citep{bur79}, where $\vec{r}$ is the position vector of the particle
with respect to the star, $\vec{v}$ its velocity vector,
$\vec{\hat{u}}=\vec{r}/r$ is the radial unit vector,
$\dot{r}=\vec{\hat{u}}\cdot\vec{v}$ is the radial velocity, and
$c$ is the speed of
light; $Q\dma{pr}$ is the dimensionless radiation pressure efficiency
coefficient averaged over the stellar spectrum.
In this expression, the velocity
independent term $(F\sigma Q\dma{pr}/c)\,\vec{\hat{u}}$ is often referred
to as the ``radiation pressure'', while the other terms constitute
the Poynting-Robertson drag (hereafter PRD) \citep{gust94}.
The radiation pressure is radial, and it is
usually described by its ratio $\bpr$ to stellar gravity
$F\dma{g}=GMm/r^2$, where $M$ is the mass of the star and $G$ the
gravitational constant. As $F\propto r^{-2}$, $\bpr$ does not depend on
$r$ for a given particle. Equation~(\ref{fullrp}) is then rewritten
as
\begin{eqnarray}
\left.\frac{\d^2\vec{r}}{\d t^2}\right|_{\mathrm{PR}}
=\frac{\bpr GM}{r^2}
\left[\left(1-\frac{\dot{r}}{c}\right)\vec{\hat{u}}-\frac{\vec{v}}{c}\right]
\, .
\label{fullrpb}
\end{eqnarray}
As $v\ll c$, the radiation pressure largely overcomes the PRD.

The long-term effect of PRD is to cause
a decrease of the orbit. Computing the rate of change of the specific
angular momentum due to PRD, and assuming that the orbit
is circular, we derive the orbital decrease as
\begin{eqnarray}
\frac{\d r}{\d t}=-\frac{2\bpr GM}{rc}\equiv -2\,\frac{\alpha}{r}\, ,
\end{eqnarray}
where $\alpha$ is a constant parameter. Numerically, we have
\begin{eqnarray}
\alpha=\beta\dma{pr}\times \pdx{6.24}{-4}\,\frac{M}{M_\odot}\,\mbox{AU}^2\,
\mbox{yr}^{-1}\, ,
\end{eqnarray}
a value already quoted by \citet{wya05}, and the characteristic PRD time-scale
writes
\begin{eqnarray*}
t\dma{PRD} = |r/(\d r/\d t)|=0.5r^2/\alpha \, .
\end{eqnarray*}
As we see below, $\bpr$ barely reaches 0.2 for submicron-sized
particles. Assuming $\bpr=0.1$ and $M=0.59\,M_\odot$ \citep{hou94},
the PRD time-scale turns out to range between $\pdx{2.7}{6}\,\mbox{yr}$
at $r=20\,$AU and $\pdx{7}{7}\,\mbox{yr}$ at 100\,AU.
As quoted by \citet{kal04}, at sufficiently large distance 
it is longer than the age of the system
\citep[$12^{+8}_{-4}$\,Myr,][]{bar99,zuc01}.

The PRD time-scale moreover needs
to be compared to the collisional time-scale, as collisions between
grains of various sizes will tend to prevent the orbital decrease
of those for which PRD is more efficient.
\begin{figure}
\centering
\includegraphics[angle=0,origin=bl,width=\columnwidth]{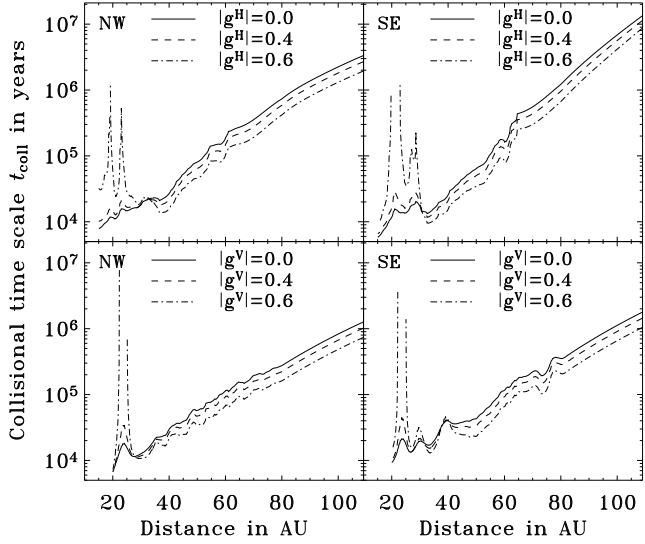}
\caption{Collisional time-scale assuming the surface densities displayed
in Fig.~\ref{midplane} and the grain size distribution
for astronomical silicates from Table~\ref{nominal}.}
\label{tcollfigure}
\end{figure}
The collision time-scale may be estimated as
\begin{eqnarray}
t\dma{coll} \simeq \frac{\sigma\dma{sca}}{\pi \langle s^2\rangle}
\times \frac{r^{3/2}}{ 2\sigma\dma{sca} \Sigma_0(r)\sqrt{G M}}
\label{tcoll}
\end{eqnarray}
where $\pi\langle s^2\rangle$ is the mean geometric cross-section of the
grains averaged over the grain size distribution \citep{bac93}. The result
is shown on Fig.~\ref{tcollfigure} for astronomical silicates. We see that
for any distance $r$, the collision time-scale is more than one order
of magnitude less that the PRD time-scale. Hence we may
safely stress that collisions dominate over PRD;
if any effect of PRD is to be investigated,
it must be averaged over the size distribution, {\it i.e.} assuming
an average $\bpr\simeq 0.01$. The corresponding PRD
time-scale turns out to be larger than the age of the star, so that
we may assume that PRD is of negligible importance
in the dynamics of the \mic\ disk. Actually, the same conclusion
is derived by \citet{wya05} for all the debris disks known today.

Assuming the Poynting-Robertson drag force on the grains is negligible
in the \mic\ disk, the equation of motion of a grain thus reads
\begin{eqnarray}
\frac{\d^2\vec{r}}{\d t^2}=-\frac{GM(1-\bpr)}{r^2}\,\vec{\hat{u}}
\, .
\end{eqnarray}
This is a classical equation of a Keplerian motion. While suffering
a radiation pressure force, a particle keeps following a Keplerian
orbit around the star, but with an effective central mass $M(1-\bpr)$.
Depending on the value of $\bpr$, that orbit may be very different from
the original orbit of the parent body. It is eccentric, shares the same
periastron as the parent body orbit, but has a very different apoastron.
At release time, the grain assumes the same orbital velocity as its
parent body. If the orbit of the parent body is circular,
energy and angular momentum preservation impose the grain orbit
to obey
\begin{eqnarray}
a'=\frac{a(1-\bpr)}{1-2\bpr}\;,\quad
e'=\frac{\bpr}{1-\bpr}\;,\quad Q'=\frac{a}{1-2\bpr}\quad.
\end{eqnarray}
Here $a$ is the semi-major axis of the original orbit, and $a'$ is 
that of the resulting dust particle orbit; $e'$ is the eccentricity
of that orbit and $Q'$ is its apoastron. If $\bpr$ is sufficiently
close to 0.5, the grain is transported at a large distance thanks
to a large $Q'$ value. If $\bpr>0.5$ it is expelled from the system.
\subsection{The spectrum of a AU\,Mic, a flare star}
\label{parsed}
\begin{figure*}
\includegraphics[angle=-90,width=\textwidth]{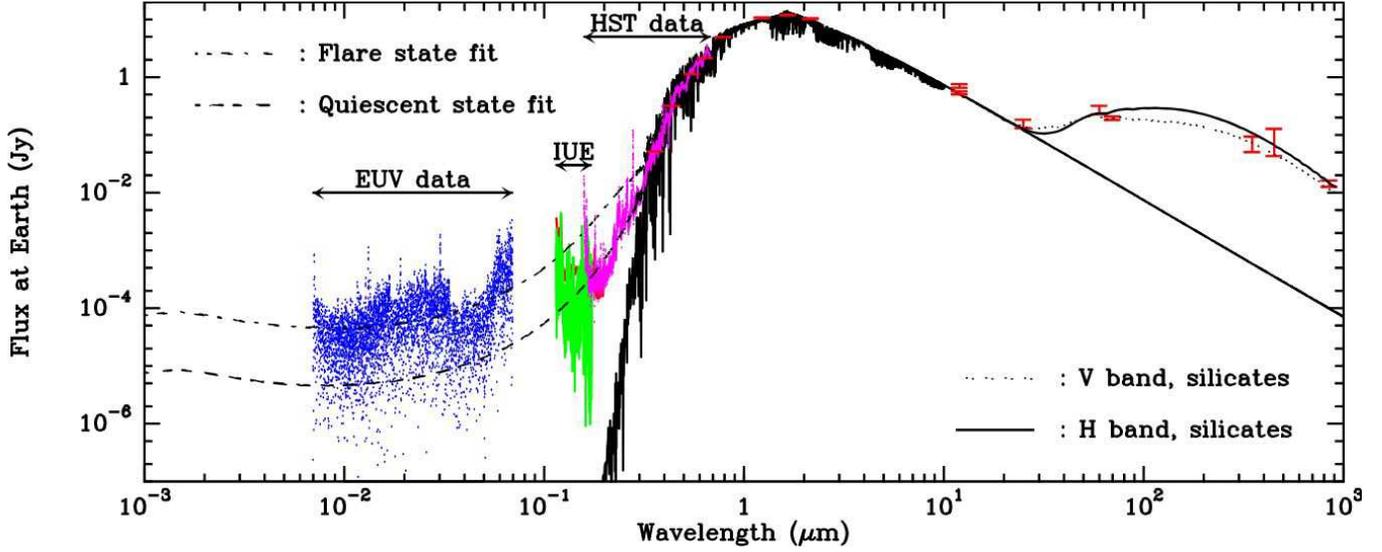}
\caption{Observed SED of the \mic\ system. The solid black curve
corresponds to the {\sc NextGen} stellar atmosphere spectrum
that fits the visible and near-infrared data.
The dashed and dashed-dotted curves are the SED fits for the quiescent
and flare states respectively (Eq.\,\ref{fitsed}).
Two fit examples to the disk SED are also overplotted 
assuming the two surface density profiles obtained
for $|g|=0.4$ (see Sect.\,\ref{mass}) and silicate grains.}
\label{sed}
\end{figure*}
The $\bpr$ ratio depends on the grain properties (size, composition)
and on the SED of the central star. 
For any given star, $\bpr$ is maximum for grains with radius $s$ in the range
0.1--1\,$\mu$m, falls off like $s^{-1}$ (geometric
regime, $Q\dma{pr}= $ constant) at larger sizes, and tends to reach
a constant for smaller grain sizes \citep[$Q\dma{pr}\propto s$, Rayleigh scattering
regime,][]{gust94}. In the solar case, $\bpr$ barely reaches 1 for
submicron-sized particles \citep{gust94,beu01}, but in the case of
classical Vega-like stars (A-type stars), $\bpr$ reaches high enough
values to enable ejection
\citep[a few micrometers in the case of \bp\ depending somehow
on the grain composition, e.g.][]{arty88,beu01,aug01}.

Contrary to those stars, \mic\ is an M-type star; hence radiation
pressure is expected to be weak \citep{kal04,liu04a,kri05}.
In order to compute it, we need to model the SED of the star.
We fit the stellar spectrum with a \textsc{NextGen}
model \citep{hau99}
with $\log(g)=4.5$\,(CGS) and an effective  temperature of 3700\,K.
This gives a stellar luminosity of $0.092$\,L$_{\odot}$. However, 
this model is not enough for describing the full SED of \mic.

\mic\ is known to frequently present X-ray and EUV flares
\citep{kun73,cully93,tsik00,mag03}. The flaring rate is
estimated by \citet{kun73} to 0.9 per hour. During these flares,
the EUV and X-ray luminosities increase by a typical factor $\sim 10$
\citep{cully93,mon96,tsik00}. We may estimate that the star spends
$\sim 10\,$\%\ of the time in the flaring state.
Even in quiescent state, the coronal activity of the star is not
negligible. \citet{tsik00} note for the X-ray and EUV luminosities
$L\dma{X}=\pdx{2.24}{29}\,\mbox{erg}\,\mbox{s}^{-1}$ and
$L\dma{EUV}=\pdx{2.9}{28}\,\mbox{erg}\,\mbox{s}^{-1}$. 
\citet{hue99} gives
$L\dma{X}=\pdx{5.5}{29}\,\mbox{erg}\,\mbox{s}^{-1}$. \citet{mit05}
reports simultaneous X-ray and UV
observations of \mic\ with XMM-Newton and give luminosities in
both domains. Their X-ray luminosity is compatible with the previous
estimate of \citet{tsik00}.
These values are
far above what should be expected from the stellar model described above.
The X-ray and EUV luminosities may contribute to the radiation pressure
felt by the dust particles, so that it is necessary to take them
into account in the stellar SED model we use. The global shape of the
X-ray and EUV spectrum of \mic\ is given in \citet{mon94};
in the UV spectral domain between 1150\,\AA\ and 2000\,\AA, IUE
data are available \citep{lands93}; Archive HST/FOS spectra cover the
spectral domain between 1600\,\AA\ and 6800\,\AA. Finally, broadband
data (U, B, V, Rc, Ic, 2MASS, IRAS, {\it Spitzer}, SHARC\,II, SCUBA) provide
additional measurements towards the infrared
\citep[][and references therein]{che05}.

Given these constraints, we fit the full spectrum of \mic\ by adding
to the \textsc{NextGen} stellar atmosphere model the following flux,
in Jansky at Earth
\begin{equation}
F(\lambda) = \exp\left[a_0+\frac{a_5}{\dy 1+\left(\frac{\ln(\lambda)-a_1}{a_2}
\right)^2}+\frac{a_3}{\lambda}-\frac{a_3a_4}{2\lambda^2}\right]\,\,\,
(\mbox{$\lambda$ in \AA})\;,
\label{fitsed}
\end{equation}
with 
\begin{eqnarray}
a_0 & = & -13.7728,\quad  
a_1\;=\; 8.9395,\qquad a_2 \;=\; 1.3165,\nonumber\\
a_3 &= & 39.9773\,\mbox{\AA},\quad
a_4\;=\;12.5702\,\mbox{\AA},\quad a_5\;=\;13.2092\:.
\end{eqnarray}
This function fits accurately the various data in the UV and EUV
domains and is compatible with the X-ray and EUV luminosities
given by \citet{tsik00}.
This holds for the quiescent state. During flares, we expect the 
flux in X-ray and UV to be higher. There is nevertheless good
evidence for synchronicity between flare events in X-ray and 
in UV \citep{tsik00,mit05}, the time-lag between the two spectral
domains being less than a few minutes \citep{mit05}. Hence we decided
to mimic flare spectra by multiplying the $F(\lambda)$ function
we add to the \textsc{NextGen} model by a fixed, average factor we estimate
to 10 \citep[see][]{tsik00}, for wavelengths shorter than
2500\,\AA\footnote{in practice, we multiplied Eq.\,(\ref{fitsed})
by $10\times(0.5-\arctan((\lambda-2500)/500)/\pi)$ with $\lambda$ in \AA}.
The result is displayed in Fig.~\ref{sed}, where we superimpose
the fits to the \textsc{NextGen} model and to the various measurements.
We see that shortwards to $\sim 0.3\,\mu$m, the resulting spectrum
is significantly above the \textsc{NextGen} model and will have
an influence on the $\bpr$ parameter felt by the various grains.

\subsection{Radiation pressure during X-ray and UV flares}
As the X-ray and UV spectrum of \mic\ is subject to fluctuations
due to flares, the $\bpr$ ratio for a given grain
may be time-variable. Hence the dynamics of the dust particles
may be more complex than described above where
we assumed a constant $\bpr$ value. As the flares are very frequent,
we cannot neglect this effect. We thus investigate now the dynamics
of solid particles with a time-variable $\bpr$.

\begin{figure}
\includegraphics[angle=-90,origin=bl,width=\columnwidth]{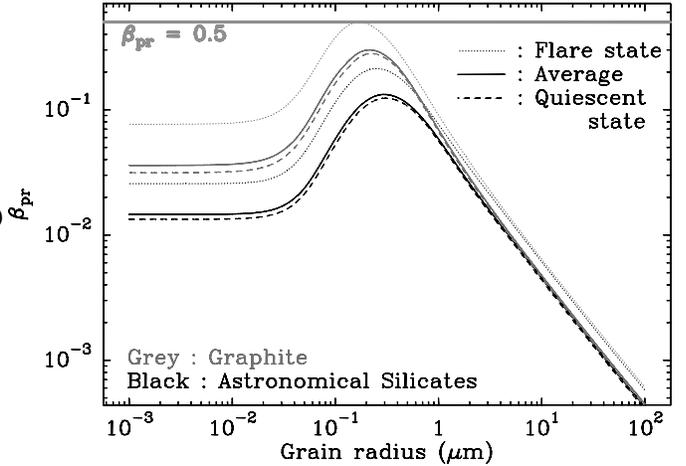}
\caption{
$\beta\dma{pr}$ ratio for astronomical silicates (black curves)
and graphite grains (grey curves)
assumed to be spherical and homogeneous, as a function of their
radius. The dashed curves correspond to the quiescent state of the
star, the dotted curves to the flare state, and the solid curves to
the temporal average.}
\label{radpressure}
\end{figure}
In a simple description, we assume that the star periodically alternates
quiescent and flare phases, respectively characterized for a given
grain by $\bprq$ and $\bprf$ ratios, with $\bprf>\bprq$. The star
is assumed to spend a fixed fraction $\gamma$ of the time in flare phases.
Strictly
speaking, the dust particle will follow bits of Keplerian orbits
alternatively characterized by dynamical masses $M(1-\bprq)$
and $M(1-\bprf)$. It can be shown (see appendix~\ref{betavar}) that
the global effect is the same as if $\bpr$ was constant, equal
to its temporal mean. The dust particle keeps following a
 Keplerian orbit around the star, but characterized by a
$\bpreff$ ratio equal to the temporal average of $\bpr$, namely
$\bpreff=(1-\gamma)\bprq + \gamma \bprf$.

Our problem reduces therefore to estimating the $\bpreff$ ratio, {\it i.e.}
the parameters $\bprq$, $\bprf$ and $\gamma$.
Figure~\ref{radpressure} shows the value of $\bpr$ as a function
of the grain radius, computed from the SED of Fig.~\ref{sed}, in
both
quiescent and flare phases, for two grain compositions and assuming
hard spheres. We also plot the $\bpreff$ average, assuming
$\gamma=0.1$. Figure~\ref{radpressure} shows that unless the grains are very
refractory, they do not suffer in quiescent phases a radiation
pressure force large enough like in the \bp\ disk to \emph{a priori} produce
an extended dust disk. For silicate grains, $\bpr$ barely reaches $0.1$
for submicron-sized grains. 
The radiation pressure is then only able to transport the grains
up to $25$\%\ of their original distance at most ($Q'=1.25a$ for $\bpr=0.1$).
In order to reach a distance of $210$\,AU \citep[maximal disk radius, ][]{kal04},
a grain produced at $40$\,AU needs $\beta\dma{pr}\simeq 0.405$, a value
never reached during quiescent phases even for refractories.
During flares conversely, the $\bpr$ ratio of graphite grains can reach
$0.5$; but the temporal mean $\bpreff$ does not. With $\gamma=0.1$
\citep[estimated from the light curves
of][]{tsik00,mag03,mit05}, $\bpreff$ remains close to $\bprq$,
and the conclusion derived above during quiescent phases still holds.
The flares help diffusing the particles slightly further out, but not
enough for explaining the brightness profile of the disk beyond
$\sim35$\,AU. Another explanation must be found.
\section{Diffusion of dust particles in the outer disk. II.~Wind Pressure Force}
\label{wind}
Another possibility for blowing dust particles into the outer disk is
the interaction with a stellar wind. Since \mic\ is a late type
star, a solar-like coronal wind is likely to arise from the star. This
is supported by the coronal activity of \mic.
\subsection{Dynamics of grains in the presence of a wind}
The force due to a flux of protons arising from the star is very similar
to the radiation force, and may be described as \citep{gust94}
\begin{eqnarray}
\left.m\,\frac{\d^2\vec{r}}{\d t^2}\right|\dma{SW}=
\Phi\dma{p}\sigma\,\frac{C\dma{D}}{2}
\left[\left(1-\frac{\dot{r}}{c}\right)\vec{\hat{u}}-\frac{\vec{v}}{c}\right]
\, ,\label{fullsw}
\end{eqnarray}
where $\Phi\dma{p}$ is the momentum density of the proton flux at the
considered stellar distance, and $C\dma{D}$ is the free molecular drag
coefficient. This equation is of course very similar to Eq.~(\ref{fullrp}),
and as for the radiation pressure case, the velocity independent term
$\Phi\dma{p}\sigma(C\dma{D}/2)\,\vec{\hat{u}}$ is dominant. Logically,
we define the dimensionless ratio $\bsw$ of that force to stellar gravity, so
that the equation of motion becomes now similar to Eq.~(\ref{fullrpb})
with $\bpr$ changed to $\bsw$. The $\bsw$ ratio has been introduced
in the solar wind case by \citet{muk82} and \citet{gust94}. We write
\begin{eqnarray}
\bsw = \frac{\Phi\dma{p}\sigma (C\dma{D}/2)}{GMm/r^2}\, .
\label{bsw}
\end{eqnarray}
The main difference between $\bpr$ and $\bsw$ is that the latter is
not necessarily independent from the stellar distance $r$. We investigate
this dependence. If $\rho\dma{sw}$ and $v\dma{sw}$ are the mass density
and velocity of the wind at distance $r$, we have
$\Phi\dma{p}=\rho\dma{sw}v\dma{sw}^2$.
The coefficient $C\dma{D}$
is given by \citet{gust94}:
\begin{eqnarray}
C\dma{D} & = & \frac{2S^2+1}{S^3\sqrt{\pi}}\,\mathrm{e}^{-S^2}
+\frac{4S^4+4S^2-1}{2S^4}\,\erf(S)\nonumber\\
&&\mbox{}+\frac{2(1-\epsilon)\sqrt{\pi}}{3S}
\sqrt{\frac{T\dma{d}}{T\dma{p}}}
\left(1+Y\dma{p}\,\frac{m\dma{s}}{m\dma{p}}\right)\;,
\label{cdeq}
\end{eqnarray}
where $S=\sqrt{m\dma{p}/(2kT\dma{p})}\,u$ is the ratio of the relative
velocity $u$ to the thermal velocity of the protons
$\sqrt{2kT\dma{p}/m\dma{p}}$ ($k$ is Boltzmann's constant),
$\epsilon$ is the fraction of re-emitted protons
at dust temperature $T\dma{d}$,
and $Y\dma{p}$ is a coefficient introduced by \citet{muk82} to account
for sputtering of molecules of mass $m\dma{s}$. In practice,
for any standard wind model, at sufficiently large distance
we have $u\simeq v\dma{sw}$ (we neglect the orbital velocity), and
for $S\gg 1$, $C\dma{D}$ is close to 2. 

If we assume spherical symmetry, then we may write the continuity equation
$\dot{M}=4\pi r^2\rho\dma{sw}v\dma{sw}$ where $\dot{M}$ is the mass
loss rate. If we assume that the grain is spherical, with radius
$s$ and density $\rho$, then the ratio $\bsw$ may be rewritten
as
\begin{eqnarray}
\bsw=\frac{3}{32\pi}\frac{\dot{M}v\dma{sw}C\dma{D}}{GM\rho\,s}\, .
\label{bsws}
\end{eqnarray}
Thus $\bsw$ may be a function of the distance to the star as
$v\dma{ws}$ and $C\dma{D}$ are not necessarily constant throughout
the disk. The stellar wind blow-out size then reads ($\bsw\ = 0.5$)
\begin{eqnarray}
s\dma{sw}=\frac{3}{16\pi}\frac{\dot{M}v\dma{sw}C\dma{D}}{GM\rho}
\,\,\,\,\,{\rm with\,\, } C\dma{D} \simeq 2.
\label{ssws}
\end{eqnarray}
\begin{figure}
\includegraphics[angle=-90,width=\columnwidth]{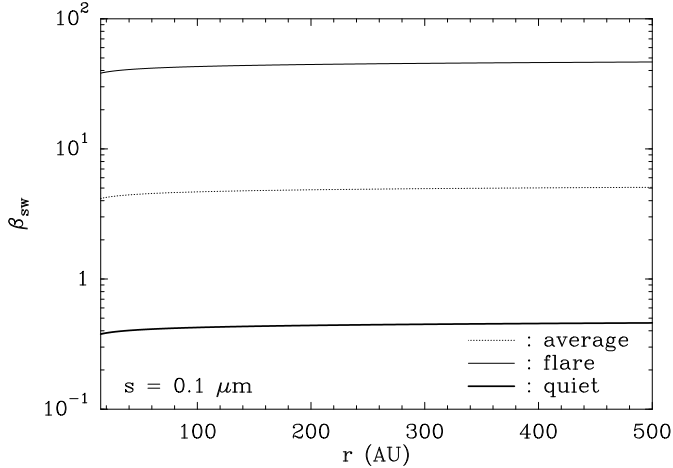}
\caption{$\bsw$ ratio as a function of the distance to the star,
computed for a spherical silicate grain with radius $s=0.1\,\mu$m, in quiescent 
(thick solid line) and flare (thin solid line) states. The dotted line
is the time-averaged $\bsw$, assuming that flares are present 10\%\
of the time ($\gamma=0.1$).}
\label{windpressure}
\end{figure}
\subsection{Wind model and Mass loss rate}
If we want to describe further we need to adopt a wind model.
The most standard model for a solar-like coronal wind is the transonic
isothermal solution by \citet{park58}. This model has the advantage of
being fully controlled by only two parameters, namely the temperature
$T\dma{p}$ and the mass loss rate $\dot{M}$, or equivalently $T\dma{p}$
and the number density $n_0$ at the base of the corona. The relation between
these parameters writes
\begin{eqnarray}
\dot{M}=4\pi R^2\,\frac{n_0m\dma{p}}{2}\,\sqrt{\frac{2kT\dma{p}}{m\dma{p}}}
\,\left(\frac{\lambda^2}{4}\mathrm{e}^{3/2-\lambda}\right)\quad,
\label{dmeq}
\end{eqnarray}
where $R$ is the radius of the star and with 
\begin{eqnarray}
\lambda=\frac{GM/R}{2kT\dma{p}/m\dma{p}}\, .
\end{eqnarray}
The velocity solution of the wind is implicitly given by
\begin{eqnarray}
w\,\mathrm{e}^{-w^2/2}=\frac{\lambda^2R^2}{4r^2}\,
\exp\left(\frac{3}{2}-\frac{\lambda R}{r}\right)\, ,
\label{parkw}
\end{eqnarray}
where $w=\sqrt{m\dma{p}/(2kT\dma{p})}\,v\dma{sw}$ is the velocity
normalized to the thermal velocity. With the above relations, we may
compute the $\bsw$ ratio as a function of the distance $r$ from the
star and of the parameters $T\dma{p}$ and $n_0$. The analytical
expression is rather complex and makes use of Lambert's $W$ function
that comes from the solution of Eq.~(\ref{parkw}), but we are
interested here into the numerical values for $\bsw$.

Our first remark is that contrary to $Q\dma{pr}$, the coupling coefficient
$C\dma{D}$ is here totally independent from the size of the grain. This
means that for spherical grains, $\bsw$ will be exactly $\propto s^{-1}$
(see Eq.~\ref{bsws}), for any grain size.
This behaviour was noted by \citet{muk82}. Hence for sufficiently
small grains, $\bsw$ may reach arbitrarily high values. 
However, \citet{muk82} show that in the case of the solar wind
characterized by $T\dma{p}\simeq 10^6$\,K and $n_0\simeq 10^9\,\mbox{cm}^{-3}$,
$\bsw$ remains several orders of magnitude less than $\bpr$ for
usual grain sizes ($\bsw$ would reach 0.5 or 1 for radii less than
$1\,$nm). As a matter of fact, in the solar environment, the wind pressure
is usually compared to Poynting-Robertson drag rather than to the full
radiation pressure, with a ratio to PRD reaching $\sim 0.3$ 
\citep{gust94}.

The situation may be drastically different with an active star like \mic.
Theoretical studies have shown that active M-dwarfs may have wind mass
loss rates typically reaching a few times
$10^{-12}\,M_\odot\,\mbox{yr}^{-1}$ \citep{war01,wood02}, i.e.
more than 100 times the solar value. Concerning \mic\ explicitly,
there are several density and temperature estimates in the coronal plasma
near the star, deduced from the observation and analysis of spectral lines
of highly ionized species. \citet{mon96}, based on the study of EUV
lines of \ion{Fe}{xxi} and \ion{Fe}{xxiv}, give a temperature
$T\dma{p}\simeq\pdx{8}{6}\,$K in quiescent phases and 2--$\pdx{5}{7}\,$K in
flare phases; they also derive electron densities (that we will take
equal to $n_0$) between $\pdx{3}{12}$ and 
$\pdx{2}{13}\,\mbox{cm}^{-3}$ during flares. \citet{pag00} find
$n_0=\pdx{6.3}{10}\,\mbox{cm}^{-3}$ in quiescent state with some
regions with $n_0=\pdx{5}{11}\,\mbox{cm}^{-3}$. Completing
that study, \citet{rob01} derive $n_0=\pdx{1.7}{12}\,\mbox{cm}^{-3}$ 
as an average for flare phases. We thus decided to assume as typical
values for quiescent state ($n_0=\pdx{6.3}{10}\,\mbox{cm}^{-3},
T\dma{p}\simeq\pdx{8}{6}\,$K), and ($n_0=\pdx{1.7}{12}\,\mbox{cm}^{-3},
T\dma{p}\simeq\pdx{3}{7}\,$K) during flares. This results via
Eq.~(\ref{dmeq}) in $\dot{M}=\pdx{9.4}{-13}\,M_\odot\,\mbox{yr}^{-1}$
in quiescent state and a peak of 
$\dot{M}=\pdx{4.9}{-11}\,M_\odot\,\mbox{yr}^{-1}$ during flares.
If we assume that flares are present 10\%\ of the time, the average
mass loss rate is $\pdx{5.7}{-12}\,M_\odot\,\mbox{yr}^{-1} \simeq
\pdx{3}{2} \dot{M}_{\odot}$, which
is in perfect agreement with the theoretical estimates quoted above.
In particular, \citet{wood02} derives a power-law relationship between
the X-ray activity of dwarfs and their mass loss rate. Applied to
\mic, this gives $\dot{M}\sim
1$--$\pdx{2}{-12}\,M_\odot\,\mbox{yr}^{-1}$
\citep{pla05}.

\subsection{Origin of the disk surface brightness profile}
We are now able to compute the $\bsw$ ratio for any grain size
at any distance from the star. This is done in Fig.~\ref{windpressure},
where $\bsw$ is plotted as a function of the stellar distance,
in quiescent and flare regimes, for a grain of radius $s=0.1\,\mu$m.
The calculation is done with $C\dma{D}$ computed from Eq.~(\ref{cdeq})
neglecting sputtering ($C\dma{D}$ turns out to remain between
2.0 and 2.1 over the whole range of stellar distances considered).
The behaviour for other grain sizes may be directly deduced from this
plot, as we strictly have $\bsw\propto s^{-1}$. First of all, we note
that there is only little variation of $\bsw$ over a wide range
of distances from the star. In fact, from Eq.~(\ref{bsws}) we see
that the dependence in $r$ of the $\bsw$ ratio is entirely
contained in the factor $v\dma{sw}C\dma{D}$. $C\dma{D}$ remains
very close to 2, and $v\dma{sw}$ is a moderately increasing function
of $r$. $\bsw$ is thus an increasing
function of $r$, but the relative increase is only 12.4\%\ between
$r=15\,$AU and $r=100\,$AU, and 8.4\%\ between $r=100\,$AU and $r=500\,$AU.
Hence considering $\bsw$ as independent from $r$, like $\bpr$, is 
an acceptable approximation. This justifies the introduction
of the $\bsw$ ratio, and  shows that the dynamics of particles
subject to wind pressure will be somewhat similar to that of particles
subject to radiation pressure. In particular, we expect that grains
will be pushed away in the outer disk as soon as $\bsw$ will approach
0.5. 

Figure~\ref{windpressure} shows that such large values of $\bsw$ can
easily be reached. For $s=0.1\,\mu$m, even in quiescent phase, $\bsw\simeq0.4$,
the exact value depending on the assumed mass loss rate.
But because of the recurrent flare episodes, 
we must consider the temporal average of $\bsw$. This is plotted
in Fig.~\ref{windpressure} as a dotted line, assuming that flares
are present 10\%\ of the time. Due to flares,
grains with $s=0.1\,\mu$m are actually blown away by the stellar
wind. As $\bsw\propto s^{-1}$, it turns out that all grains 
with $s\la 1\mu$m are able to diffuse into the outer disk, and
even be ejected from the system. Actually the exact behaviour at large
distance may depend on the location of the heliopause of the stellar
wind, but this is not expected closer to a few hundreds of AU.

The fact that $\bsw$ is almost constant
over several orders of magnitude of stellar distances, and that 
$\bsw$ is able to reach sufficiently large values, show that
particles subject to wind pressure will behave almost as if they
were subject to a comparable radiation pressure. This explains why
the surface brightness profile along the disk midplane
($\propto r^{-4\cdots-5}$) resembles that of the 
\bp\ disk, the latter being due to the effects of the intense
radiation pressure \citep{lec96,aug01,the05}. 

From this study we conclude that in the \mic\ system,
the stellar wind pressure is expected to play almost the same
role as the radiation pressure in other debris disks around
earlier type stars like \bp. Interestingly, \citet{rob05} also came
to the conclusion that the short lifetime of the circumstellar
gas in the \mic\ system can only be explained by dissipation mechanisms
other than photo-evaporation, including a possible stellar wind.
\citet{pla05} also proposed that the dynamics of the grains
in the inner AU\,Mic system could be controlled by the drag
component of the stellar wind. Given that the collision
time-scale we find within $\sim30\,$AU compares to the
stellar wind drag time-scale inferred by \citet{pla05}
(time-scale rescaled assuming $\dot{M}\simeq\pdx{3}{2} \dot{M}_{\odot}$),
the drag force may play a significant role in explaining the
disk clearing in the inner regions.
Interestingly, we find that the role of the flares,
neglected in previous studies, is significant.
%
%
%%%%%%%%%%%%%%%%%%%%%%%%%%%%%%%%%%%%%%%%%%%%%%%%%%%%%%%%%%%%%%%%%%%%%%%%%%
\section{Summary and Conclusion}
\label{concl}
We have calculated the surface density profiles of the edge-on debris
disk around the M-type star \mic\ by performing a direct inversion of
two brightness profiles: the H-band profiles obtained by \citet{liu04a}
with the Keck adaptive optics system and the HST/ACS V-band
profiles from \citet{kri05}. We have solved the integral equation
that links the scattered light surface brightness profile along the
disk midplane to the disk surface density. The solution to this
equation only depends on the light scattering phase function
characterized by the asymmetry parameter $|g|$, which is
constrained to be smaller than $0.6$.
From the analysis of the H-band observations, we find that the dust
distribution peaks around $35\,$AU, close to the location of the power
law breakup in the surface brightness profile.
The surface density profiles inferred from the inversion of the visible
images are quite different, and this difference may partly come from
dust properties varying throughout the disk midplane.
Even if the exact azimuthal profile of
the disk cannot be  unambiguously derived, the disk turns out to be obviously
a non-symmetrical ring, as the two NE and SW branches are not identical. 

From the analysis of the disk colors measured in the visible with the
HST/ACS instrument \citep{kri05}, we constrain the grain size
distribution to be steeper than the classical $s^{-3.5}$ differential
size distribution. Grains as small as $0.1\,\mu$m in radius are in
addition required to explain the observations when astronomical
silicates are assumed.  We infer a total mass of silicates grains smaller
than $1\,$mm equal to $3.5$--$\pdx{8}{-4}\,M_{\oplus}$. This dust mass
compares the mass of the Ceres asteroid in the Solar System, but it is $14$
to $30$ times smaller than the dust mass inferred from millimeter observations
by \citet{liu04b}. It is also $9$ to $20$ times smaller than the mass we infer
from the disk SED fitting. This difference cannot be explained by the disk
optical thickness since the disk is optically thin in all directions
and at all wavelengths ($\tau_{\parallel} < 0.04$ at visible
wavelengths).  Possible explanations for this discrepancy include the
fact that the simple power law size distribution that we find from
the interpretation of the visible observations can probably not be
extrapolated all the way to millimeter grain sizes \citep[see e.g. the
discussion in][]{the03}. It may also be possible that assuming hard
spheres to calculate the grain optical properties is not appropriate
if the grains have complex structures.

We then explored whether the similarities between the surface
brightness profiles of the \mic\ and \bp\ disks are coincidental or if
they could result from similar dynamical processes.
The global properties of the dust distribution in the \bp\ disk
are very well explained by the role of radiation pressure. In this
context, the breakup in surface brightness (observed at 120\,AU)
corresponds to the outer edge of the planetesimal disk that produces
the dust particles \citep{aug01}. A very similar picture may
apply to the \mic\ disk but at a distance of $\sim 35\,$AU.
The main difference lies in the dust diffusion mechanism.
The radiation pressure from \mic\  appears inefficient
to diffuse dust into the outer disk
even when the X-ray and UV flares are taken into account.
Conversely, the corpuscular pressure from a stellar wind is
very likely to be much more efficient. In fact, in the \mic\ 
environment, the wind pressure turns out to play an almost
identical role as the radiation pressure in the \bp\ disk,
with in both cases, a constant (or almost constant) ratio to
stellar gravity. This explains why the surface brightness  profiles
of the two disks are similar, despite different generating
mechanisms.

This comparison between debris disks about early and late-type stars
would be generic if the additional role of flares was not
significant in the case of \mic.
The flares contribute to considerably enhance the effect
of the stellar wind, and consequently to transport in the outer disk
many additional particles that would otherwise remain close
to their parent bodies. Debris disks may be common around late-type
stars, but an intense and frequent coronal activity may help
to make them sufficiently extended to be detectable.
%
%
%
%%%%%%%%%%%%%%%%%%%%%%%%%%%%%%%%%%%%%%%%%%%%%%%%%%%%%%%%%%%%%%%%%%%%
%%%%%%%%%%%%%%%%%%%%%%%%%%%%%%%%%%%%%%%%%%%%%%%%%%%%%%%%%%%%%%%%%%%%
\begin{acknowledgements}
We thank M. Liu, P. Th\'ebault, the FOST (Grenoble) and the AstroChem (Leiden) team
members, as well as the referee, P. Kalas, for helpful comments and suggestions.
This work was partly supported by the European Community's
Human Potential Programme under contract HPRN-CT-2002-00308,
PLANETS. %
\end{acknowledgements}
%
%
%
%
%%%%%%%%%%%%%%%%%%%%%%%%%%%%%%%%%%%%%%%%%%%%%%%%%%%%%%%%%%%%%%%%%%%%%%
\appendix
%
%---------------------------------------------------------------
\section{Numerical scheme to invert Eq.\,(\ref{Abel3})}
\label{algo}
The product integration method is a simple but efficient and generally 
numerically stable technique that can be used to invert an integral
equation. This Appendix describes the main steps towards the
implementation of this method in order to solve Eq.\,(\ref{Abel3}).
We refer to Sect.~\ref{inversion} and \ref{vprofile} for the notations
used in this appendix.

We expand Eq.\,(\ref{Abel3}) as a sum
of $n$ integrals over ranges of equal size $h$
\begin{eqnarray*}
S(y) = \sum_{j=0}^{n-1} \int^{r\dma{max}-jh}_{r\dma{max}-(j+1)h}
\sigma\dma{sca}\,\Sigma_0(r)\,
\frac{A(r,y)}{r\sqrt{r^2-y^2}} \, dr
\end{eqnarray*}
where $y=r\dma{max}-nh$ and
\begin{eqnarray*}
A(r,y)=\Phi^* D_z^{-1}(r,\theta) \,\Theta(r,\theta)\,(f(\theta)+f(\pi-\theta)) \,\, .
\end{eqnarray*}
We note $r_j=r\dma{max} - jh$ which gives $r_n=y$ and $r_0=r\dma{max}$.
$S(y)$ and $A(r,y)$ are supposed to be two known functions of $r$ and/or $y$.

We assume that for $r$ within the
range $r_{j+1} \leq r \leq r_{j}$, $\sigma\dma{sca}\,\Sigma_0(r)\,A(r,y)$
can be approximated, to first order, by its value 
at the middle of the interval of size $h$ (product integral method), namely
at the position $r=r_{j+\frac{1}{2}}$. The smaller is $h$, the better is
the approximation. The surface brightness profile $S(y)=S(r_n)$ thus reads
\begin{eqnarray*}
S(r_n) & \simeq & \sum_{j=0}^{n-1}\,\sigma\dma{sca}
\Sigma_0(r_{j+\frac{1}{2}})A(r_{j+\frac{1}{2}},r_n) I(r_{j+1},r_j,r_n) \\
\mathrm{where\,\,\,} &  & I(r_{j+1},r_j,r_n) = \int^{r_j}_{r_{j+1}}
\,\frac{1}{r\sqrt{r^2-r_n^2}} \, dr \\
\mathrm{and\,\,\,} & & \int\,\frac{1}{r\sqrt{r^2-r_n^2}} \, dr = 
-\frac{1}{r_n}
\arctan{\left(\frac{r_n}{\sqrt{r^2-r_n^2}}\right)} \, .
\end{eqnarray*}
$h$ being small compared to the size of the disk,
taking $n=1$ in the above expression of $S(r_n)$ leads to a first
crude estimate of $\sigma\dma{sca}\,\Sigma_0(r)$ close the outer
edge $r\dma{max}$ of the disk
\begin{eqnarray*}
\sigma\dma{sca}\,\Sigma_0(r_{\frac{1}{2}}) \simeq 
\frac{S(r_1)}{A(r_{\frac{1}{2}},r_n) I(r_{1},r_0,r_n)} \,\,\, .
\label{Rx0.5}
\end{eqnarray*}
For $n\ge 2$, we can write
\begin{eqnarray*}
S(r_n) \simeq
\sum_{j=0}^{n-2}\sigma\dma{sca}\,\Sigma_0(r_{j+\frac{1}{2}})
A(r_{j+\frac{1}{2}},r_n)
I(r_{j+1},r_j,r_n) + \\ \sigma\dma{sca}\,\Sigma_0(r_{n-\frac{1}{2}})
A(r_{n-\frac{1}{2}},r_n)
I(r_{n},r_{n-1},r_n)
\end{eqnarray*}
and then
\begin{eqnarray*}
\sigma\dma{sca}\,\Sigma_0(r_{n-\frac{1}{2}}) \simeq 
\frac{S(r_n)}{A(r_{n-\frac{1}{2}},r_n) I(r_{n},r_{n-1},r_n)} \hspace{1.5cm}\\
- \frac{\sum_{j=0}^{n-2}\sigma\dma{sca}\,\Sigma_0(r_{j+\frac{1}{2}})
A(r_{j+\frac{1}{2}},r_n)
I(r_{j+1},r_j,r_n)}{A(r_{n-\frac{1}{2}},r_n)
I(r_{n},r_{n-1},r_n)} \,\,\, .
\label{Rxnm0.5}
\end{eqnarray*}
With this simple algorithm, we can directly invert Eq.\,(\ref{Abel3})
to get the product of the surface number density $\Sigma_0(r)$ by the
scattering cross section $\sigma\dma{sca}$ as a function of the distance
to the star. $\sigma\dma{sca}\,\Sigma_0$ is reconstructed from the outer
disk edge ($r\dma{max}$) down to the closest possible distance.
At each distance $r$ from the star, $\sigma\dma{sca}\,\Sigma_0(r)$
is thus calculated using the values of $\sigma\dma{sca}\,\Sigma_0$
previously estimated at distances larger than $r$.
We checked that with this simple algorithm we could reproduced
the analytic results obtained by \citet{nak90} for an edge-on disk with a
decreasing power-law surface brightness profile.
%
%
%---------------------------------------------------------------
\section{Accounting for non-axisymmetric disks: the $\Theta(r,\theta)$
smoothing function}
\label{theta}
The inversion of the NW and SE surface brightness profiles
do not depend on the position angle on the skyplane at which they
has been measured. To simplify the discussion, and as far as the inversion
is concerned, we will thus always assume that the brightness profile 
is measured along the $y$--axis and orientated toward the
positive $y$ direction (see Fig.~\ref{sketch}).

If $h$ (as defined in Appendix~\ref{algo}) could be infinitely small,
the inversion of Eq.\,(\ref{Abel3}) would
lead to the reconstruction of the surface density of the disk
precisely along the $y$--axis, namely at a polar angle $\theta=\pi/2$. But 
because  $\sigma\dma{sca}\Sigma_0(r)$ is estimated at
the middle of an interval of finite size $h$, the surface density
is calculated at polar angles $\theta_0$ that slightly deviates from
$\pi/2$ and with $\lim \theta_0 = \pi/2$ when $h\rightarrow 0$.
The smoothing function $\Theta(r,\theta)$ introduced in Sect.~\ref{vprofile}
in order to account for non-axisymmetric disks reads
\begin{eqnarray}
\Theta\dma{SE}(r,\theta) & = & \frac{1-p}{2} \times
\left(\frac{\theta}{\theta_0}\right)^{\delta} + \frac{1+p}{2}
\label{thetase}
\end{eqnarray}
for the SE side of the disk, and
\begin{eqnarray}
\Theta\dma{NW}(r,\theta) & = & \frac{1-p^{-1}}{2}\times
\left(\frac{\theta}{\theta_0}\right)^{\delta} + \frac{1+p^{-1}}{2}
\label{thetanw}
\end{eqnarray}
for the NW side with $0\leq \theta \leq \pi/2$ and
where $p$ is given by
\begin{eqnarray}
p =\left[\sigma\dma{sca}\Sigma_0(r)\right]\dma{NW} /
\left[\sigma\dma{sca}\Sigma_0(r)\right]\dma{SE}
\end{eqnarray}

\section{Dynamics of particles with time-variable radiation pressure}
\label{betavar}
We investigate here the dynamics of particles with a time-variable $\bpr$.
We assume that the star periodically alternates
quiescent and flare phases, respectively characterized for a given
grain by $\bprq$ and $\bprf$ ratios, with $\bprf>\bprq$. The star
is assumed to spend a fixed fraction $\gamma$ of the time in flare phases.
Strictly
speaking, the grain follows bits of Keplerian orbits
alternatively characterized by dynamical masses $M(1-\bprq)$
and $M(1-\bprf)$. What is the global effect ? First, we should
note that this problem cannot be treated by a classical perturbation
method typical for celestial mechanics, as the ``perturbation''
(the flares effect) overcomes the main Keplerian forces. 

Let us consider a simple episode, characterized by a first quiescent
phase ($\bpr=\bprq$), a flare ($\bpr=\bprf$),
and a second quiescent phase ($\bpr=\bprq$). During each of these phases
the grain follows a different Keplerian orbit. Due to the flare,
the Keplerian orbit during the first quiescent phase differs from that
during the second quiescent phase. We want here to compute the 
orbital change. 

First, we note that during the whole episode, the angular momentum
remains unchanged. Even if $\bpr$ is time-variable, the resulting
acceleration keeps pointing towards the central star, keeping the angular
momentum constant. Hence the orbital plane of the motion remains
unchanged and so the quantity $a(1-e^2)$ during the successive
quiescent phases. Second the orbital energy $\delta E$ change between the
two quiescent phases may be computed easily, using the continuity of
the velocity during the whole episode. After a simple algebra,
we derive
\begin{eqnarray}
\delta E = GM(\bprf-\bprq)\,\frac{r_2-r_1}{r_1r_2}\, ,
\label{dde}
\end{eqnarray}
where $r_1$ and $r_2$ are the distances of the particle to the star
at the beginning of the flare and at the end respectively.
Now, the important remark is that the period of recurrence of
the flares (a few days at most) at far less than the orbital period
of the particle (hundreds of years at 35--40\,AU). The distance
change during the whole episode may be regarded as infinitesimal.

Let us now consider the time span between the beginning of a flare
and the beginning of the next flare. During this time span, we first
have a flare that generates a (quiescent) orbital energy change
described by Eq.~(\ref{dde}), and a quiescent phase that keeps is
constant up to the next flare. The time-span considered may be 
regarded as infinitesimal compared to the orbital period, so that
the motion of the particle during this phase is nearly uniform.
If $\d r$ is the infinitesimal distance change during this phase,
we should have $r_2-r_1\simeq\gamma\d r$ in Eq.~(\ref{dde}). 
The orbital energy change within $\d r$ reads then
\begin{eqnarray}
\d E = GM(\bprf-\bprq)\,\frac{\gamma\,d r}{r^2}\, .
\end{eqnarray}
Integrating this equation, we derive the macroscopic energy change
that corresponds to the stellar distance change from $r$ to $r'$~:
\begin{eqnarray}
\Delta E = GM\gamma(\bprf-\bprq)\,\left(\frac{1}{r}-\frac{1}{r'}\right)
\, .
\end{eqnarray}
This is equivalent to say that the quantity
\begin{eqnarray}
E\dma{eff} = E+\frac{GM\gamma(\bprf-\bprq)}{r}
\end{eqnarray}
Given the expression of the energy $E$, we derive
\begin{eqnarray}
E\dma{eff} & = & \frac{1}{2}v^2-\frac{GM(1-\bprq)}{r}
+\frac{GM\gamma(\bprf-\bprq)}{r}\nonumber\\
 & = & \frac{1}{2}v^2-\frac{GM(1-\bpreff)}{r}
\end{eqnarray}
where
\begin{eqnarray}
\bpreff=(1-\gamma)\bprq+\gamma\bprf\, .
\label{beff}
\end{eqnarray}
$E\dma{eff}$ is just the energy corresponding to a Keplerian orbit
with central mass
$M(1-\bpreff)$, i.e. to a particle feeling a $\bpr=\bpreff$ ratio
from the star. Combined to the preservation of angular momentum,
we see that on average, the dust particle will keep following a
 Keplerian orbit around the star, but characterized by a $\bpr=\bpreff$
ratio. We checked this result numerically with a dedicated integrator.

$\bpreff$ is only the temporal average of the $\bpr$ ratio between flare
and quiescent phases. Finally, the net results is that whenever
the radiation pressure is time-variable, the grain behaves as if
it was feeling a $\bpr$ ratio from the star that is just the temporal
average of the variable $\bpr$. We should nevertheless not forget that
this is valid if the recurrence frequency of the flares largely overcomes
the orbital frequency of the orbit. Significantly different outcomes
should be expected for instance if there was a resonance between the
mean motion of the orbit and the recurrence frequency of flares, which
is not the case here.

%%%%%%%%%%%%%%%%%%%%%%%%%%%%%%%%%%%%%%
%
%
%%%%%%%%%%%%%%%%%%%%%%%%%%%%%%%%%%%%%%%%%%%%%%%%%%%%%%%%%%%%%%%%%%%%%%
%

%
\end{document}